\documentclass[aps, prx, reprint, groupedaddress,superscriptaddress,amsmath,amssymb,floatfix]{revtex4-2}
\usepackage{graphicx}
\usepackage{dcolumn}
\usepackage{bm}
\usepackage{makecell}
\usepackage{caption}
\usepackage{placeins}
\usepackage{braket}
\usepackage{color}
\usepackage{comment}
\usepackage{sidecap}
\usepackage{multirow}
\usepackage{todonotes}
\usepackage{bm}
\newcommand{\rev}[1]{{\color{black} #1}}
\newcommand{\wavenumber}[1]{#1~\text{cm}^{-1}}
\usepackage{hyperref}

\renewcommand{\thesubsection}{\thesection.\arabic{subsection}}

\renewcommand{\thetable}{\arabic{table}}

\begin{document}

\title{Accurate Machine Learning Interatomic Potentials for Polyacene Molecular Crystals: Application to Single Molecule Host-Guest Systems}

\author{Burak Gurlek}
\thanks{These two authors contributed equally}
\affiliation{Max Planck Institute for the Structure and Dynamics of Matter and Center for Free-Electron Laser Science, Luruper Chaussee 149, 22761 Hamburg, Germany}

\author{Shubham Sharma}
\thanks{These two authors contributed equally}
\affiliation{Max Planck Institute for the Structure and Dynamics of Matter and Center for Free-Electron Laser Science, Luruper Chaussee 149, 22761 Hamburg, Germany} 

\author{Paolo Lazzaroni}
\affiliation{Max Planck Institute for the Structure and Dynamics of Matter and Center for Free-Electron Laser Science, Luruper Chaussee 149, 22761 Hamburg, Germany}

\author{Angel Rubio}
\affiliation{Max Planck Institute for the Structure and Dynamics of Matter and Center for Free-Electron Laser Science, Luruper Chaussee 149, 22761 Hamburg, Germany}
\affiliation{Initiative for Computational Catalysis (ICC), The Flatiron Institute,
162 Fifth Avenue, New York, New York 10010, USA}

\author{Mariana Rossi}
\email{mariana.rossi@mpsd.mpg.de}
\affiliation{Max Planck Institute for the Structure and Dynamics of Matter and Center for Free-Electron Laser Science, Luruper Chaussee 149, 22761 Hamburg, Germany}
\begin{abstract}
\begin{center}\textbf{Abstract}\end{center}
Emerging machine learning interatomic potentials (MLIPs) offer a promising solution for large-scale accurate material simulations, but stringent tests related to the description of vibrational dynamics in molecular crystals remain scarce. Here, we develop a general MLIP by leveraging the graph neural network-based MACE architecture and active-learning strategies to accurately capture vibrational dynamics across a range of polyacene-based molecular crystals, namely naphthalene, anthracene, tetracene and pentacene. Through careful error propagation, we show that these potentials are accurate and enable the study of anharmonic vibrational features, vibrational lifetimes, and vibrational coupling. In particular, we investigate large-scale host-guest systems based on these molecular crystals, showing the capacity of molecular-dynamics-based techniques to explain and quantify vibrational coupling between host and guest nuclear motion. Our results establish a framework for understanding vibrational signatures in large-scale complex molecular systems and thus represent an important step for engineering vibrational interactions in molecular environments.
\end{abstract}

\maketitle

\section{Introduction}
Organic molecular crystals, characterized by their long-range order and rich intermolecular interactions, are crucial in diverse applications, ranging from pharmaceuticals to electronics, and hold significant potential for emerging technologies, such as photovoltaics~\cite{Congreve2013} and quantum information systems~\cite{toninelli2021}.  
While these applications primarily rely on the underlying electronic properties of these systems, molecular vibrations, encompassing both inter- and intramolecular modes, are equally important due to their role in determining the crystal structure and the pronounced electron-phonon coupling which is often observed~\cite{Alvertis2022, Devos1998, Vukmirovi2012, Kato2002, neefarxiv2024}. 

Specifically, molecular vibrations and their anharmonic couplings play a pivotal role in determining the thermodynamic stability of crystal polymorphs~\cite{Nyman2015,Rossi2016,KrynskiRossinpj2021,KapilEngelPNAS2022, HojaSciAdv2019}, in enhancing or hindering charge transport by modulating carrier mobility through dynamic intermolecular coupling~\cite{Coropceanu2007,Chang2022,neefarxiv2024}, in facilitating rapid singlet fission to improve solar cell efficiency~\cite{Congreve2013,seilersciadv2021,neefpssa2024}, and even in offering a potential usage as quantum memory elements~\cite{Gurlek2021}. For example, polycyclic aromatic hydrocarbons embedded in large-bandgap host materials are being explored as single-photon sources, nonlinear quantum optical elements, and nanoscale sensors~\cite{toninelli2021}, as they exhibit narrow optical transitions at cryogenic temperatures, allowing highly coherent light-matter interactions~\cite{basche-2008-book}. However, previous studies have predominantly focused on the electronic transitions, leaving the rich internal structures arising from vibrational, and spin degrees of freedom largely unexplored~\cite{gurlek2024}. 
Despite their undeniable importance, accurately modeling vibrational dynamics that are affected by anharmonic mode-coupling and long-range van der Waals interactions is hampered by the computational complexity of such simulations, which makes them prohibitively expensive with traditional first-principles methods such as density-functional theory (DFT).

Machine learning interatomic potentials (MLIPs) hold great promise in addressing the challenges associated with large-scale and long-time simulations of complex material systems, offering high computational efficiency without compromising accuracy~\cite{Behler2007,Bartk2010,jinnouchi2019fly,Batzner2022,Batatia2022mace, Deringer2019,Smith2019,Ko2023}. Recent developments in active-learning strategies and in strategies for sampling diverse atomic environments have further enabled the  construction of smaller representative training datasets~\cite{cersonsky2021improving,schran2020committee,karabin2020entropy,allotey2021entropy} that deliver good training accuracy. These methods and training strategies have allowed the generalization of these potentials through the proposition of foundational models that can perform well for a large variety of systems, including those not represented in their training set~\cite{Chen2022,batatia2023foundation,kovacs2023maceoff,merchant2023scaling,allen2024learning,deng2023chgnet,yang2024}. 

Despite the success of MLIPs in modeling solids~\cite{Schmidt2019, Verdi2021}, solid-liquid interfaces~\cite{Wan2024}, and chemical reactions~\cite{Yang2024-reactive}, assessments of their performance and reliability in describing vibrational dynamics in molecular crystals remain limited. Existing studies have mostly focused on inorganic and covalently-bonded systems~\cite{Bartk2018,Loew2024, bandi2024, lee2024, monserrat2020liquid, Verdi2021}. Molecular crystals are particularly challenging due to their soft vibrational modes, which are impacted by van der Waals (vdW) intermolecular interactions. The full long-range and non-local character of vdW interactions is normally not captured by MLIP architectures based on local atomic environments, which calls the accuracy of these architectures into question. 

\rev{In this work, we develop MLIPs capable of accurately capturing harmonic and anharmonic vibrational dynamics in polyacene molecular crystals. Starting from naphthalene as a model system, we systematically develop MLIPs and assess their predictive accuracy across the polyacene series to pentacene, demonstrating that the potentials can generalize to larger acenes and to previously unseen host-guest configurations. We further assess the reliability of these MLIPs by showing how errors on forces propagate to phonon frequencies and anharmonic vibrational densities of states (VDOS). These measures allow for a rigorous predictive confidence of these quantities and for devising active learning targets for vibrational properties.

The extrapolative capabilities of these MLIPs  to predict vibrational properties of host-guest systems, in particular pentacene molecules embedded in a naphthalene crystal host, provide important new insight for vibrational control. We present a clear assessment of host and guest vibrational mode assignment when anharmonic correlations play an important role, thus providing a foundation for the study of vibrational coherence and decoherence processes in molecular host-guest systems.}

\section{Results}
\subsection{Performance of VASP and MACE Machine-Learning Potentials Based on Active Learning}\label{sec:VASPvsMACE}

To investigate the accuracy of MLIPs for modeling vibrational dynamics in molecular crystals, we compare the performance of MLIPs obtained from the VASP and the MACE machine-learning algorithms for the naphthalene crystal. We train both models using a dataset generated through the active learning strategy in VASP~\cite{jinnouchi2019fly}, which employs on-the-fly sampling of structures from a molecular dynamics (MD) trajectory based on uncertainties in the predicted energy, forces and stresses, utilizing Bayesian regression.
%%%%%%%%%%%%%%%%%%%%%%%
\begin{table}[h] % Table environment
    \centering % Center the table
    \caption{The training errors of VASP and MACE MLIPs
for the naphthalene molecular crystal, based on a single-temperature training at $295$~K. The errors are presented as root mean square errors (RMSE) for energies and forces.} 
    \label{tab:errors_vaspmace} % Add label for referencing
    \begin{tabular}{ ccc }
         & Energy (meV/atom) & Force (meV/\AA) \\ 
        \hline
        
        VASP   & 0.1  & 23.7  \\ 
        MACE   & 0.1  & 10.5   \\ 
        \hline
    \end{tabular}
\end{table}
%%%%%%%%%%%%%%%%%%%%%%%%%

We created the training dataset by running  MD trajectories with VASP at $295$ K for a $1\times 2\times 2$ naphthalene supercell. During the MD run, we monitored the changes in the number of training structures and the forces root mean square error (RMSE) of the model. We stopped the active learning process when these metrics showed negligible change, resulting in $1402$ structures in the training dataset (see details in Methods). These structures were then used to train a MACE equivariant message-passing machine learning potential~\cite{Batatia2022mace}. The MACE MLIP training is continued until the RMSE of forces and energy on the validation set converged across different epochs~(see details in Methods). To assess the stability of resulting MLIPs, we conducted a $1$~ns NVT-MD run on a $4 \times 4 \times 4$ naphthalene supercell, which showed no signs of instability. 

We compare the performance of the two MLIPs using the RMSE for energy and forces on the training dataset, as shown in Table~\ref{tab:errors_vaspmace}. These metrics indicate that the MACE MLIP outperforms
the VASP MLIP, particularly in predicting atomic forces, which is reflected on the accuracy with which it can predict harmonic phonon frequencies, as shown in the Supplementary Note~1. We note that the MACE MLIP predicts energies more accurately for structures at temperatures close to its training temperature of $295$~K 
(see relevant discussions in Supplementary Note~1). Overall, the improved performance of the MACE model on the transferred dataset could be due to its longer effective interaction range~\cite{niblet2024} or to the higher effective body order achieved with the message-passing procedure. The VASP model we trained contains up to 9-body order in the kernels and cutoffs of 8~\AA\ for radial descriptors and 5~\AA\ for angular descriptors, while the MACE model trained in this work goes up to 13 body-order and 12~\AA\ effective cutoff. 
In addition, as shown in Supplementary Note~2, the performance of the VASP model on a transferred dataset obtained through a different active-learning strategy is relatively poor, indicating a limitation in data transferability similar to those observed in other MLIP architectures~\cite{niblet2024}.\rev{We note that this transferred dataset, covering multiple temperatures, is the same dataset later used for the MACE potential, which is described in the next paragraph.}

%%%%%%%%%%%%%%%%%%%%%%%%%%%%%
\begin{table}[h] % Table environment
    \centering % Center the table
    \caption{The training and validation errors of VASP MLIP-multi and MACE MLIP-committee for the naphthalene molecular crystal, based on training data at $80$~K, $120$~K, $150$~K, $220$~K and $295$~K, accounting for thermal lattice expansion on each dataset. The errors are presented as RMSE for energies and forces.} 
    \label{tab:errors_bestvaspmace} % Add label for referencing
    \begin{tabular}{ p{1.1cm} >{\centering\arraybackslash}p{1.6cm} >{\centering\arraybackslash}p{1.6cm} >{\centering\arraybackslash}p{1.6cm} >{\centering\arraybackslash}p{1.6cm} }
         & \multicolumn{2}{c}{Energy (meV/atom)} & \multicolumn{2}{c}{Force (meV/\AA)} \\ 
        \hline
         & Training & Validation & Training & Validation \\
         \hline
        VASP   & 0.1 &0.2    & 20.9 &14.6    \\ 
        MACE   & 0.1 &0.1  & 4.3 &4.4   \\ 
        \hline
    \end{tabular}
\end{table}
%%%%%%%%%%%%%%%%%%%%%%%%%%%%%%%%%%%%
\begin{figure}[ht]
    \centering
    \includegraphics[width=\columnwidth]{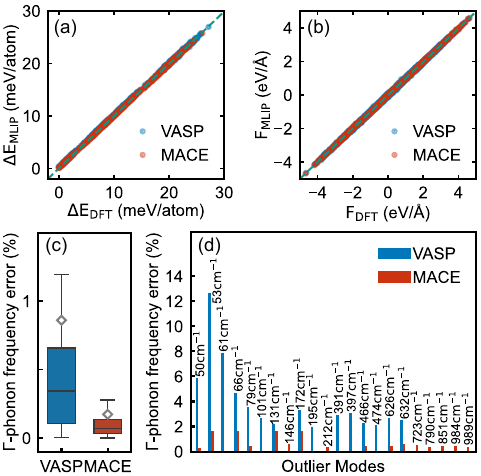}
        \caption{\textbf{Comparison of VASP MLIP-multi and MACE MLIP-committee for the naphthalene molecular crystal}.~\textbf{a} Correlation plot of relative energies, $\Delta E_\text{DFT(MLIP)}=E_\text{DFT(MLIP)}-E_\text{DFT}^\text{min}$, and \textbf{(b)} forces predicted by DFT and MLIPs.~\textbf{c} Error box plot of harmonic phonons ($\Gamma$-point) obtained with MLIPs.~\textbf{d} Outliers identified from the box plot in \textbf{c}. The wavenumbers of the modes correspond to those of the associated modes in corresponding VASP reference DFT calculations. }
         \label{fig:bestvaspvsmace}
\end{figure}

We next examine the impact of different active-learning strategies on the overall performance of MLIPs. To achieve this, we construct a new VASP MLIP by sequentially sampling MD trajectories at temperatures of $295$~K, $220$~K, $150$~K, $120$~K, and $80$~K using VASP’s active learning algorithm, explained in Ref.~\cite{jinnouchi2019fly}. The final training dataset consists of $1168$ naphthalene structures, distributed as $940$, $22$, $145$, $14$, and $47$ structures for each respective sampling temperature. We will refer to the resulting MLIP as VASP MLIP-multi for the remainder of the text. To construct the MACE MLIP, we employ a committee-based active learning strategy~\cite{schran2020committee} explained in Methods. We simultaneously trained a committee of eight MACE MLIPs on an initial dataset of 100 structures, selected from a pool using farthest point sampling (FPS)~\cite{fps_eldar, cersonsky2021improving} (see details in Methods, Computational Details).
At each active-learning iteration, $25$ structures were added to the training dataset based on energy uncertainty, resulting in a total of $450$ naphthalene crystal structures. The number of structures selected at each respective temperature were $290$, $71$ $30$, $33$,$26$. We refer to the resulting model as MACE MLIP-committee in the following discussions.

In Table~\ref{tab:errors_bestvaspmace}, we present a comparison of training and test errors for the VASP MLIP-multi and MACE MLIP-committee models, evaluated on an independent test dataset composed of $2100$ naphthalene crystal structures (see details in Methods, Computational Details). The VASP MLIP-multi shows training errors comparable to those of the initial VASP MLIP (see Table~\ref{tab:errors_vaspmace}) and may exhibit a sample-bias issue within the training dataset, as the test errors on forces are slightly smaller than the training errors. The MACE MLIP-committee outperforms the earlier MACE MLIP by achieving lower errors in predicting atomic forces. The close agreement between training and test errors for MACE MLIP-committee indicates that it is not overfitted, nor does it suffer from sample bias. Furthermore, in Fig.~\ref{fig:bestvaspvsmace}(a, b), we compare the MLIP and DFT-predicted energies and force components of each structure in the test set, showing good predictive performance for both VASP MLIP-multi and MACE MLIP-committee.

In Fig.~\ref{fig:bestvaspvsmace}(c), we compare the performance of the new MLIPs in predicting $\Gamma$-point phonon frequencies. The MACE MLIP-committee outperforms the VASP MLIP-multi, with mean percentage (absolute) frequency errors of $0.17\%$ ($\wavenumber{0.98}$) and non-outlier maxima of $0.27\%$, corresponding to only $\wavenumber{2.88}$. Error distributions for outlier modes, shown in Fig.\ref{fig:bestvaspvsmace}(d), reveal that VASP MLIP-multi struggles with accurately predicting intermolecular vibrations. The MACE MLIP-committee achieves absolute frequency errors below $\wavenumber{3.5}$ with mean frequency errors of $\wavenumber{0.48}$ for intermolecular, $\wavenumber{1.03}$ for intramolecular and $\wavenumber{1.39}$ C-H stretching modes, surpassing the predictive capabilities of other MLIPs [see Supplementary Fig.~4(b)]. Besides, the overall performance of MACE MLIP-committee is significantly improved compared to MACE MLIP, demonstrating the effectiveness of the committee-based active learning approach in capturing diverse atomic configurations. 

These results demonstrate that both VASP and MACE MLIPs can achieve comparable accuracy in predicting the vibrational properties of naphthalene molecular crystals. However, when combined with a committee-based active-learning algorithm, the MACE model yields the best accuracy. Therefore, in the remainder of this work, we will focus on the MACE MLIP.

\subsection{Committee Uncertainty Propagation for Vibrational Properties}\label{sec:Uncertainty}

Determining the reliability and confidence of any MLIP prediction relies on being able to calculate uncertainties for directly predicted quantities, as well as for quantities derived from such predictions. Various methodologies have been proposed for quantifying uncertainties in MLIP predictions of energies and forces~\cite{zhu2023fast, heid2024spatially, venturi2020bayesian, musil2019fast, imbalzano2021uncertainty, kellner2024uncertainty, schran2020committee, jinnouchi2019fly}, and for propagating these uncertainties to static observables~\cite{imbalzano2021uncertainty, kellner2024uncertainty,bauer2024roadmap}. Quantifying uncertainties in dynamical (time-dependent) observables is also critical in the context where anharmonic vibrational couplings, vibrational lifetimes, and transport coefficients are derived from molecular dynamics simulations~\cite{Kubo1957}. In these simulations, the time-evolved atomic motion is governed by forces derived from the MLIP's potential energy surface. Errors in force predictions, especially for underrepresented or rare atomic configurations in the training dataset, will propagate to these observables. Therefore, we use our committee model to propagate uncertainties in MLIP predictions to the harmonic phonon frequencies and anharmonic vibrational density of states (VDOS), as discussed below.

\begin{figure}[t]
\centering
    \includegraphics[width=\columnwidth]{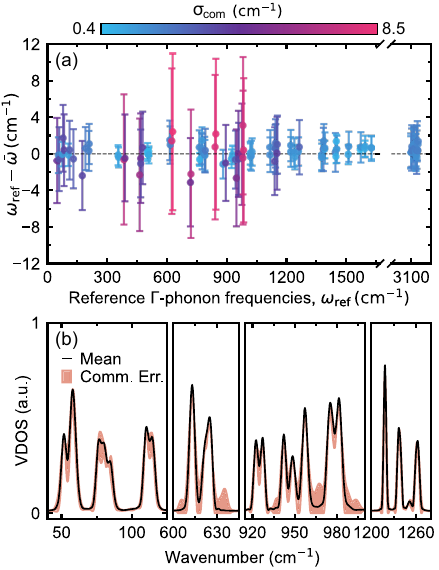}
        \caption{\textbf{Uncertainty propagation of phonon frequencies and VDOS in a naphthalene crystal at 80~K.}~\textbf{a} The uncertainity estimations for the $\Gamma$-point phonon frequencies. The mean committee predictions ($\bar{\omega}$) are compared to reference DFT calculations ($\omega_\text{ref}$), with error bars color-coded to represent the standard deviation among committee members.~\textbf{b} Uncertainty estimations for the VDOS. The red shaded area and the black curve represent the committee error calculated by  Eq.~\eqref{eq:com_error} and the VDOS calculated by using the committee mean force, respectively. The committee error is only shown when it is larger than the statistical uncertainty.}
        \label{fig:PhononError_Propagation}
\end{figure}

Employing the MACE MLIP-committee model, we first calculate the committee uncertainty for $\Gamma$-point harmonic phonon frequencies \rev{as detailed in Methods}. In Figure~\ref{fig:PhononError_Propagation}~(a), we show the distribution of relative errors between committee predictions and reference DFT calculations for the phonon spectrum. Overall, we conclude that the propagated uncertainty shows a good prediction capability of the real error across the whole frequency range. The strong correlation between the uncertainty and the real error suggests that committee uncertainty can serve as a reliable measure of error, especially when computing the true error is computationally expensive. 
The largest uncertainties appear in the region between 600 and 1000 cm$^{-1}$. This region is dominated by modes that involve the in-plane and out-of-plane deformations of the fused benzene rings, as well as the in-plane and out-of-plane bending of CH groups (see Supplementary Fig.~6).~\rev{As shown in Supplementary Fig.~7, sampling the displacements along these high-uncertainty modes in a committee-based active-learning procedure substantially improves both the predictive accuracy and uncertainty estimations of the corresponding normal modes, without the need for brute-force molecular dynamics}. We do not observe instances of the uncertainty underestimating the real error for this model.

Next, we tackle a much harder problem, related to the propagation of the uncertainty in committee-MLIP predictions to the anharmonic VDOS, detailed in Methods. To illustrate the procedure for the naphthalene crystal, we perform molecular dynamics equilibration runs at $80~\text{K}$ for each committee member, propagating the dynamics using rescaled forces computed via Eq.~\eqref{eq:scaled-force}. Afterwards, we perform 100 NVE simulation runs of 20 ps using the same forces for the propagation and calculate the VDOS$_{\bm{F}_i}$ corresponding to a given committee member (Eq.~\eqref{eq:vdos}).
We call VDOS$_{\overline{\bm{F}}}$ the VDOS computed using the committee's mean forces $\overline{\bm{F}}$, which is the standard quantity computed when using committee models. Statistical uncertainties $\sigma_\text{stat}$ for VDOS$_{\overline{\bm{F}}}$ and each VDOS$_{\bm{F}_i}$ are obtained from the block average of 100 NVE runs, and $\sigma_\text{com}$ is determined by removing $\sigma_\text{stat}$ from the total uncertainty over all committee members.

An issue with this procedure is that due to the non-linear dependence of the VDOS on forces, the VDOS obtained from averaging the predictions of all committees, $\overline{\text{VDOS}_{\bm{F}_i}}$, is not equal to $\text{VDOS}_{\overline{\bm{F}}}$. However, as we show in Supplementary Fig.~8, both spectra are quite similar. We can formally only calculate the uncertainty on $\overline{\text{VDOS}_{\bm{F}_i}}$ with the procedure we follow in this work, and we therefore  take this uncertainty as a proxy for the uncertainty in $\text{VDOS}_{\overline{\bm{F}}}$. The standard error on the mean of the spectra is reported.

In Fig.~\ref{fig:PhononError_Propagation}~(b), we show these results in distinct frequency regions corresponding to intermolecular and intramolecular vibrations. We confirm the robustness of the statistical sampling across the entire frequency range and find that the statistical error is overall small (see Supplementary Fig.~9). Furthermore, the committee error is generally comparable in magnitude to the statistical error, indicating that the variability among committee members is on par with statistical fluctuations. However, around $\wavenumber{600}$ and within the range $\wavenumber{900-1000}$, the committee error shows that different committees would predict peak positions differently, leading to larger errors related to peak positions. Interestingly, these regions correlate with the regions of largest uncertainties on harmonic phonon frequencies shown in Fig.~\ref{fig:PhononError_Propagation}~(a). 

Such a careful uncertainty quantification shows that the MACE-MLIP committee model can deliver accurate harmonic and anharmonic vibrational properties of the naphthalene molecular crystal. It also defines the limits within which peak positions and widths can be interpreted, based on committee uncertainty. This approach separates model uncertainty from statistical noise, which mainly affects spectral intensities, allowing estimation of MLIP-related errors. We find that the uncertainties on harmonic modes correlate with the errors in the anharmonic dynamical vibrational spectra, making them useful for assessing VDOS-prediction reliability, at least at lower temperatures.

\subsection{Generalizing Machine-Learning Potentials for Polyacene Molecular Crystals}\label{sec:Generalization}
\begin{figure*}[t]
\centering
%\begin{figure}[ht]
%\begin{center}
\includegraphics[width=18cm]{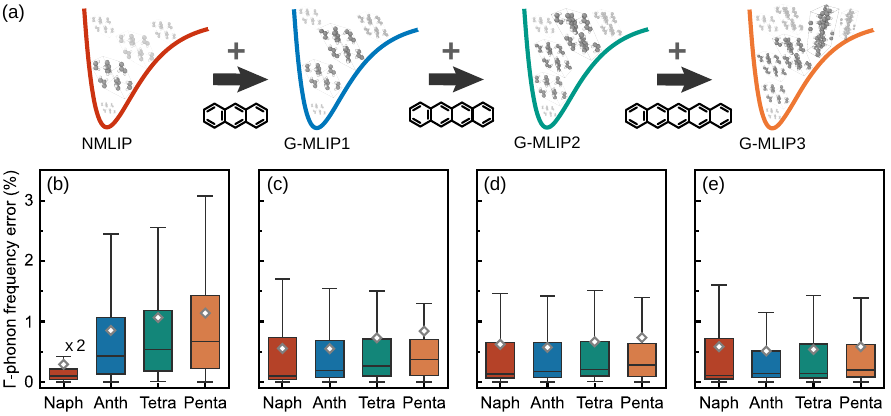}
\caption{\textbf{Generalized MLIPs for polyacene crystals: development and validation.}~\rev{\textbf{a} Sketch showing the active-learning scheme used to create generalized MLIPs.~\textbf{b-e} Error box plots for the $\Gamma$-phonon frequencies of naphthalene (Naph), anthracene (Anth), tetracene (Tetra) and pentacene (Penta) molecular crystals, predicted by N-MLIP, G-MLIP~$1-3$, respectively.}}
\label{fig:Fig_MultiMLFF}
%\end{center}
\end{figure*}

Next, we investigate the capability of the MACE MLIP to generalize across acene-based molecular crystals for the prediction of vibrational dynamics. We choose to train our own general potentials, instead of using a foundational model such as MACE-OFF~\cite{kovacs2023maceoff},
because in this work an important goal is to carefully benchmark the quality of potentials for vibrational properties. Therefore, training a model from scratch with data coming from codes we can fully control, including a uniform definition of the DFT functional, basis sets and other numerical settings, is paramount. In addition, the committee-based uncertainty quantification  requires the ability to generate and subsample datasets, which we can create specifically for this purpose.

We employ a systematic active-learning strategy based on the MACE MLIP-committee developed in the previous section. The MLIP is progressively generalized by sequentially incorporating molecular crystal structures with an increasing number of fused benzene rings, i.e., anthracene, tetracene, and pentacene molecular crystals, as schematically shown in Fig.~\ref{fig:Fig_MultiMLFF}~(a). \rev{The similarity of the crystals across the acene-based series  makes it more likely that the generalization of the MLIP will be successful.} The total pool of training structures for the each molecular crystal are reported in Table~\ref{tab:dataset_pool}. After each active-learning step, we assess the performance of the resulting MLIP in predicting harmonic $\Gamma$-point phonon frequencies for naphthalene, anthracene, tetracene, and pentacene molecular crystals.

We begin by evaluating the generalization capabilities of the MACE MLIP-committee developed in Section~\ref{sec:VASPvsMACE} without including any additional data. Hereafter, we label this MLIP as N-MLIP. To evaluate the stability of N-MLIP, we performed $2$~ns long NVT MD runs at $295$ K on $1 \times 2 \times 2$ supercells of anthracene, tetracene and pentacene molecular crystals, observing  no signs of instabilities. We then analyzed the performance of N-MLIP in predicting phonon frequencies as shown in Fig.~\ref{fig:Fig_MultiMLFF}~(b). Compared to its performance on the naphthalene molecular crystal, the accuracy of N-MLIP is nearly five to ten times lower for anthracene, tetracene and pentacene molecular crystals (see also Supplementary Fig.~10 for outlier modes and  Supplementary Table~1 for absolute errors). This limitation is particularly pronounced in the maximum absolute frequency errors reaching up to $\wavenumber{40}$ in the case of the tetracene molecular crystal. Despite the stability of the N-MLIP, its predictive accuracy for vibrational dynamics within the acene family is very limited.

To address this limitation and improve the generalization of the potential, anthracene molecular crystal structures are added to the training dataset using the active-learning strategy. Following five consecutive active-learning steps, $125$ anthracene structures were incorporated into the dataset, resulting in an updated MLIP, denoted as G-MLIP1. Stability tests confirm that G-MLIP1 is robust across all studied molecular crystals~(see details in Methods, Committee-based active learning strategy). As illustrated in Fig.~\ref{fig:Fig_MultiMLFF}(c), G-MLIP1 achieves nearly twofold improvements in phonon frequency predictions for anthracene, tetracene, and pentacene crystals (see also Supplementary Table~1). However, this generalization comes at the cost of reduced accuracy for naphthalene, attributed to a trade-off between specificity and broader applicability across the acene family. Nevertheless, the inclusion of anthracene configurations significantly enhances the extrapolative capacity of the MLIP.  

We then examine the impact of expanding the training dataset with additional tetracene molecular structures. G-MLIP2, built by incorporating 150 primitive-cell structures of tetracene through six successive active-learning steps, demonstrates stability across all tested systems. While it shows a slight improvement in overall performance, as illustrated in Fig.~\ref{fig:Fig_MultiMLFF}(d), it notably reduces maximum phonon frequency errors by nearly $\wavenumber{10}$ for tetracene and pentacene molecular crystals, highlighting its enhanced generalization to more complex molecular environments~(see Supplementary Table~1).

Building on this observation, we further expanded the training dataset by incorporating 125 pentacene structures obtained through five consecutive active-learning steps. The resulting G-MLIP3 remains stable across all acene molecular crystals and demonstrates slightly improved performance over earlier models~[see Fig.~\ref{fig:Fig_MultiMLFF}(e)]. Importantly, it achieves a consistent average error of approximately $\wavenumber{2.8}$ across all systems, highlighting its ability to generalize effectively to diverse molecular environments while avoiding overfitting to any specific system~(see Supplementary Table~1). This robust generalization is further evident in its threefold reduction of the maximum phonon frequency error for the pentacene molecular crystal.

A closer examination of the errors associated with G-MLIPs reveals distinct performance trends for intermolecular and intramolecular vibrational modes throughout the molecular crystals investigated. G-MLIP3 stands out with the lowest mean errors, consistently below $\wavenumber{3.0}$~(Supplementary Table~1), effectively capturing the key physical characteristics of both high-frequency intramolecular modes and low-frequency intermolecular modes. This good performance is particularly prominent for intramolecular vibrations in the range of $\wavenumber{1000-1600}$, while for the lower-frequency range of $\wavenumber{100-250}$, all the G-MLIPs exhibit similar error performance~(see Supplementary Fig.~11). As a generalization test for G-MLIP3, we further assessed its performance on the vibrations of crystal polymorphs of tetracene and pentacene, as shown in Supplementary Fig.~13. The potential is as accurate for different polymorphs as it is for the polymorph it was trained on.~\rev{In addition, we analyze whether the committee uncertainties remain predictive of the potential’s accuracy by propagating the uncertainties to harmonic phonon frequencies of naphthalene, anthracene, tetracene, and pentacene molecular crystals, following Section~\ref{sec:Uncertainty}. As shown in Supplementary Fig.~12, G-MLIP3 generally exhibits smaller uncertainties for the naphthalene crystal compared to MACE MLIP-multi [see Fig.~\ref{fig:PhononError_Propagation}(a)], despite its lower predictive performance. This indicates that the committee model is slightly overconfident and underestimates the actual error, particularly in the 150–1000 cm$^{-1}$ range. However, all errors we quantify are already very small.}

These results demonstrate that the generalized MLIPs derived from the multi-acene active-learning strategy exhibit good performance and robust MD stability. The inclusion of larger molecular structures into the training dataset results in a clear accuracy improvement for vibrational properties,  highlighting the benefits of diversifying the dataset with closely related molecular structures. Next, we use G-MLIP3 to study the vibrational dynamics in acene-based host-guest systems, which include atomic environments not represented during training, testing the model's ability to extrapolate to new, unseen structures.

\subsection{Vibrational Correlations in a Host-Guest System}\label{sec:hostguest}

An attractive property of single-molecule host-guest systems is that, once engineered, the vibrational levels in these systems could feature long coherence times that exceed those of the electronic transitions, potentially paving the way for the realization of quantum memories and efficient optomechanical interactions~\cite{Gurlek2021}. Therefore, to fully harness the potential of molecular host-guest systems, it is important to develop a deeper understanding of their vibrational dynamics — a challenge that traditional \textit{ab-initio} methods struggle to address. Here, we validate the G-MLIP3 potential developed in the previous section for host-guest systems and apply it to explore their vibrational properties, at this point keeping a classical description of anharmonic nuclear motion.
%%%%%%%%%%%
\begin{figure}[t]
\centering
\includegraphics[width=\columnwidth]{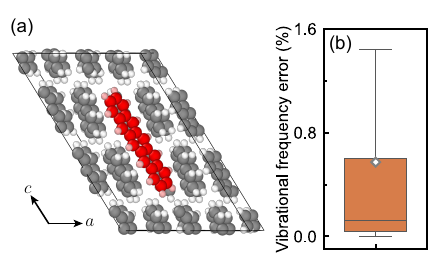}
\caption{\textbf{Extrapolative performance of G-MLIP3 in host–guest systems.}~\textbf{a} Crystal structure of pentacene-doped $2 \times 2 \times 3$ naphthalene molecular crystal, with $a$ and $c$ representing the crystal axes in monoclinic symmetry. Two naphthalene molecules along the front-facing plane are removed for illustration purposes.~\textbf{b} Box plot showing the prediction errors for vibrational frequencies of the host-guest system.}
\label{fig:Fig_HG_phonon}
\end{figure}
%%%%%%%%%%%

We investigate the pentacene-doped naphthalene molecular crystal due to its compatibility with the generalized MLIP and its relevance in both experimental and theoretical studies~\cite{Kummer1996,steiner2024}. A schematic visualization of this system is shown in Fig.~\ref{fig:Fig_HG_phonon}(a), where a pentacene molecule replaces two naphthalene molecules in a $2\times 2\times 3$ naphthalene supercell. Using G-MLIP3 alongside reference DFT calculations, we first relax the host-guest molecular crystal and evaluate the accuracy of G-MLIP3. The insertion energy error for the guest molecule in the host crystal is found to be $0.1$~meV per atom, consistent with the test set error (see Table~\ref{tab:errors_bestvaspmace}). Furthermore, we compute the vibrational frequencies of the host-guest system and benchmark these results against reference DFT calculations. As illustrated in Fig~\ref{fig:Fig_HG_phonon}(b), the mean vibrational frequency error is less than $1\%$, and its overall performance is on par with G-MLIP3 (see Fig.~\ref{fig:Fig_MultiMLFF}), demonstrating the ability of MLIPs to generalize to unseen atomic configurations, with absolute errors consistently below $\wavenumber{15}$ (see Supplementary Fig.~14). 

\rev{As shown in Supplementary Note~3, we find that G-MLIP3 rapidly loses accuracy under isotropic cell expansion due to its inability to capture long-range van der Waals (vdW) interactions, whereas NMLIP retains accuracy for small expansions when predicting properties of the naphthalene crystal. When we analytically include vdW interactions in the potential, long-range effects can be accurately described even for very large lattice expansions in both generalized and specialized MACE MLIPs (see Supplementary Note~3). In the following simulations, we fix the lattice constants to the experimental values of the $295$~K naphthalene structure.}

In order to gauge the reliability of the vibrational property predictions of G-MLIP3 on the host-guest system, we analyse the propagated uncertainties on the harmonic phonons. \rev{In Supplementary Fig.~15, we show that even for the significantly larger $4\times 4\times 5$ pentacene-naphthalene supercell containing $8640$ vibrational modes, the committee uncertainty remains below $\wavenumber{11}$ across the entire vibrational spectrum.}

%%%%%%%%%%%%%%%%%%%%
\begin{figure*}[ht]
\centering
\includegraphics[width=17cm]{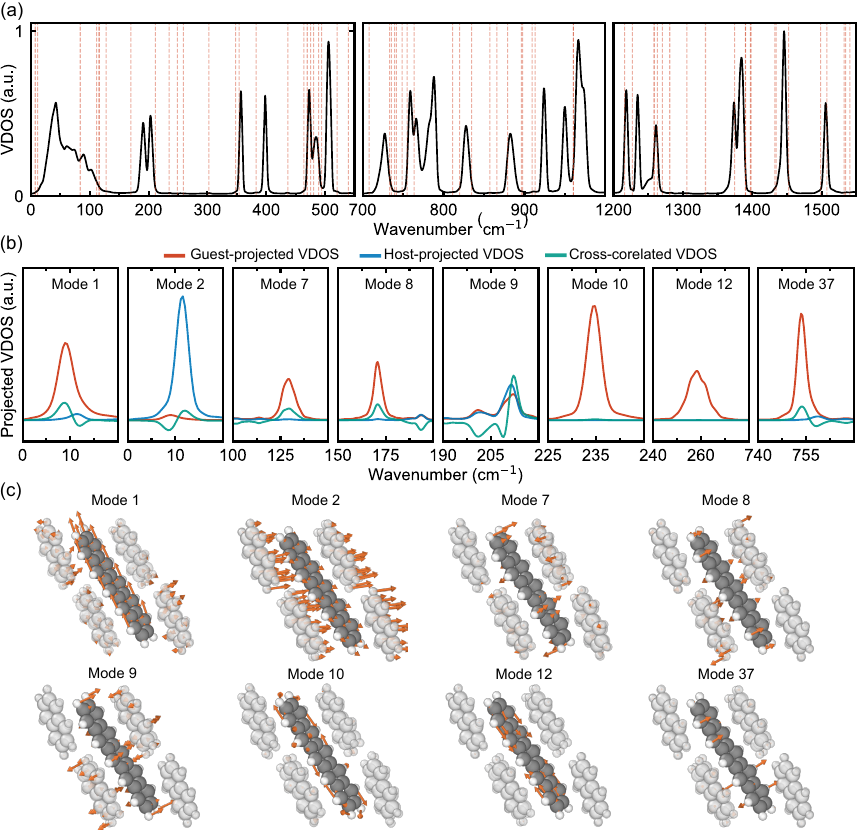}
\caption{\textbf{Vibrational dynamics in host–guest systems captured by G-MLIP3.}~\textbf{a} VDOS for the $4 \times 4 \times 5$ host-guest system. The red dashed lines indicate the frequencies of vibrational modes where the atomic displacements of the guest molecule dominate over those of the host molecules.~\textbf{b} Normal-mode-projected VDOS for the host-guest system, with the red, blue and green lines representing guest-projected, host-projected and cross-correlated VDOSs, respectively.~\textbf{c} Illustrations of the normal modes of the host-guest system discussed in (b) in Cartesian space, along the $ac$ crystal plane, showing only the most relevant naphthalene molecules from the $4\times 4\times 5$ supercell. For clarity, the displacement vectors of the host and guest molecules are scaled by a factor of $3:1$, except for modes $2$ and $9$, which are scaled by $1:1$. For more detailed illustrations, see Supplementary Fig.~19}
\label{fig:Fig_HG_VDOS}
\end{figure*}
%%%%%%%%%%%%%%%%%%%%

After confirming the reliability of the generalized MLIP for the host-guest system, we apply it to the analysis of its vibrational dynamics. The vibrational landscape of combined host-guest systems can be rationalized by considering the vibrations of the isolated host and isolated guest systems. The host crystal exhibits a continuum of intermolecular vibrational modes (often named phonons) up to $\wavenumber{150}$, and a series of intramolecular vibrational bands, while the guest molecule has discrete intramolecular vibrational modes (often named vibrons) starting at frequencies around $\wavenumber{30}$, depending on its size~\cite{gurlek2024}. When these two systems are combined, their vibrational modes hybridize, resulting in guest-like, host-like, and mixed modes. The guest-like modes in the phonon spectral region are referred to as \emph{pseudolocal modes}, characterized by a decaying vibrational amplitude away from the guest molecule~\cite{gurlek2024}. These pseudolocal modes can play a crucial role in explaining the temperature-dependent dephasing of electronic transitions~\cite{skinner-1988}. 

In high-resolution vibronic spectroscopy of single-molecule host-guest systems, optical scattering is observed at frequencies corresponding to transitions to the continuum of host phonons, guest-like vibrations, and pseudolocal modes~\cite{Myers1994, Kummer1996, Zirkelbach-2021}. Transitions to high-frequency host-like modes above the phonon cut-off frequency of~$\wavenumber{150}$, are negligible due to weak electron-vibration coupling strengths~\cite{nazir-2016}. Despite previous studies on specific vibrational modes in these systems~\cite{Fleischhauer1992, Bordat-2000, Deperasiska2017, Zirkelbach-2021}, a comprehensive understanding and characterization of these modes and their correlations remains largely unexplored.

We calculate the VDOS of a $4 \times 4\times 5$ pentacene-doped naphthalene molecular crystal (2880 atoms) at 100~K (see details in Methods, Computational Details). The VDOS in Fig.\ref{fig:Fig_HG_VDOS}(a) is analyzed in three spectral regions where prominent vibronic transitions have been observed~\cite{Zirkelbach-2021}. To rationalize these features, we also calculate harmonic phonons and identify $98$ harmonic normal modes with significant atomic displacements in the guest molecules.
%on average, relative to the host. 
These are highlighted as potential guest-like modes in Fig.~\ref{fig:Fig_HG_VDOS}(a) with red dashed lines. It is worth noting that a visual comparison of harmonic and anharmonic spectra of this host-guest system at $100$~K does not immediately show any clear signs of anharmonicity below $\wavenumber{800}$, see Supplementary Fig.~18.

The continuum of phonon modes is clearly observed below $\wavenumber{150}$, where the spectrum peaks around $\wavenumber{45}$, corresponding to the low group velocities of naphthalene optical phonon modes. This peak gradually decays toward the phonon cut-off frequency, in agreement with experimental observations~\cite{Kummer1996,Zirkelbach-2021}. In this region, we identify seven modes with the potential to represent pseudolocal modes. The remainder of the spectrum above $\wavenumber{150}$ exhibits vibrational bands originating from intermolecular and intramolecular vibrational modes of naphthalene molecules (isolated bands) and mixed vibrational modes (bands coinciding with guest-like vibrational modes). We also observe minor peaks between these bands, coinciding with the red dashed lines, which may correspond to guest-like vibrations, e.g., around $\wavenumber{230}$ and $\wavenumber{430}$. Note that, due to the large number of degrees of freedom in the host molecules, the contributions from the guest molecules are not easily discernible in the total VDOS, but can be more clearly seen in the projected VDOS of the guest molecule (see Supplementary Fig.~17). 

To gain a deeper understanding of the vibrational mode structure, we project the VDOS onto the 98 potential guest-like normal-modes of the host-guest system, and separate the contributions from the host and the guest degrees of freedom as outlined in Methods. This projection yields: (i) a partial projected VDOS arising solely from the guest molecular motions; (ii) a partial projected VDOS accounting for vibrational contributions solely from the host molecules; and (iii) the cross-correlated VDOS, which contains the cross contributions of host and guest vibrations in the normal mode basis. The cross-correlated VDOS term reflects  potential hybridization effects in the vibrational spectrum. 

\rev{In Fig.~\ref{fig:Fig_HG_VDOS}~(b), we present an example of this projection for several guest-like modes across various spectral regions, including pseudolocal vibrational modes (below $\wavenumber{150}$) and the most relevant guest-like modes, illustrating their mode displacements in Fig.~\ref{fig:Fig_HG_VDOS}~(c). Among the seven candidate pseudolocal modes [red-dashed lines below $\wavenumber{150}$ in Fig.~\ref{fig:Fig_HG_VDOS}~(a)], only modes $1$ and $7$ exhibit strong localization on the guest molecule with small contributions from the host atoms as evidenced by the cross-correlated VDOS. These modes correspond to hindered backbone bending and wave-like out-of-plane deformations, with corresponding frequencies of $\wavenumber{36}$ and $\wavenumber{100}$, respectively, for pentacene in vacuum. Mode $1$ is coupled to the nearby host-like mode $2$ in an anti-correlated manner, likely arising from the strong hybridization of the same backbone bending mode of pentacene with host phonons [see Fig.~\ref{fig:Fig_HG_VDOS}(c)]. This anti-correlation behavior suggests anharmonic coupling between these two modes. Interestingly, the mode at $\wavenumber{130}$ lies at the phonon band edge and exhibits a non-Lorentzian lineshape, consistent with experimental observations~\cite{Zirkelbach-2021} that report sharp spectral features in this region.
 
Moreover, modes 10 and 12, which involve in-plane torsional and stretching motions, do not overlap with any host spectral features and remain largely localized on the guest molecule.  
In contrast, mode 8, which exhibits out-of-plane torsional motion, shows a mixed character and couples to host intermolecular vibrations near $\wavenumber{200}$, despite also lacking any direct spectral overlap with host vibrations . This highlights that mode coupling is influenced not only by spectral overlap but also by nonlinear effects in the potential energy landscape. Such anharmonic couplings could explain discrepancies between earlier spectroscopic mode assignments and ab initio results~\cite{Zirkelbach-2021}.
%Note that despite the strong localization of guest modes, they may exhibit anharmonic effects, as observed for mode 12. 

Finally, modes 9 and 37 exemplify the effect of spectral overlap with host bands around $\wavenumber{200}$ and $\wavenumber{760}$, respectively. Mode 9, involving wave-like out-of-plane deformations, exhibits mixed character with strong anharmonic coupling to nearby host modes. On the other hand, mode 37, featuring rocking deformations of the guest molecule, retains its guest character and shows only weak anharmonic interactions despite the energy overlap with the host modes.

Using the VDOS normal-mode projections, we classified 80 vibrational modes as guest-dominated (with an integrated guest-projected VDOS greater than $90\%$ of total VDOS) and $11$ mixed modes (including $2$ pseudolocal modes) that exhibit strong guest character, based on the relative contributions from both guest and host components. Additionally, by fitting the normal-mode projected VDOS with a Lorentzian lineshape, we obtained the lifetimes of the guest-dominated vibrational modes, which averaged 3.5 ps (see Supplementary Fig.~20). This analysis is only expected to be accurate for low-frequency modes, below $\approx$~$\wavenumber{300}$, for which employing classical dynamics for the nuclei at 100~K is expected to be valid. Nevertheless, these lifetimes are shorter for modes with frequencies below twice the phonon cut-off frequency of the naphthalene molecular crystal, and longer above the cutoff, as previously discussed by Dlott \textit{et al.}~\cite{Dlott-1989}.}

\section{Discussion}\label{sec:Conclusion}

Our work demonstrates the potential of MLIPs in accurately modeling the vibrational dynamics of polyacene molecular crystals, with a particular focus on their ability to generalize across related chemical spaces and predict vibrational properties and vibrational correlations of host-guest systems. Recent developments in MLIPs have led to the so-called foundational models, also based on the MACE architecture, which are transferable across large variety of systems. In particular, it is important to put the results presented in this paper within the context of MACE-OFF~\cite{kovacs2023maceoff}, which is a foundational model targeted at the description of organic molecules and molecular condensed-phase systems. 

The potential we developed achieves errors comparable to the large models of MACE-OFF for energies and forces while keeping the maximum angular momentum of equivariant features equal to the small model (i.e., only invariant features of $L=0$). By leveraging active-learning strategies, an excellent accuracy on energies, forces and vibrational properties was achieved for naphthalene, anthracene, tetracene and pentacene crystals with only a few hundred DFT calculations in total.

With this strategy, we could propagate and quantify errors in dynamical quantities such as anharmonic vibrational spectra. Our incorporation of uncertainty propagation into vibrational properties enables a quantitative assessment of the MLIP predictions of these quantities. We show that the error quantification for the harmonic spectra, when done carefully, serves as a good proxy for identifying the vibrational motions for which the potential is least accurate also in anharmonic spectra derived from molecular-dynamics. Even though we expect this relationship to degrade with increasing temperature, we propose that harmonic phonon uncertainties are incorporated on active learning strategies in order to improve the potential for anharmonic vibrational properties, as it is computationally quite challenging to calculate the uncertainties in the latter case. Indeed, we expect to use this strategy on top of general foundational models to fine-tune them for more reliable anharmonic vibrational analysis.

We also note that while the general MLIPs we developed for the polyacene crystals do not contain explicit long-range interactions, they show good accuracy also for low-frequency phonon modes where intermolecular forces dominated by van der Waals interactions play a dominant role. \rev{ 
However, as demonstrated in Supplementary Note~3, explicit inclusion of vdW corrections is essential to accurately capture lattice expansion and contraction at larger scales.}   

The training strategy we have followed produced potentials that are also able to very accurately extrapolate to host-guest systems not contained in the training set, in particular regarding their anharmonic vibrational properties. We have presented an analysis of a pentacene molecule embedded into a naphthalene crystal matrix, unraveling guest-host vibrational mode correlations in exquisite detail and providing a measure of anharmonic coupling that goes beyond a simple analysis of mode-energy overlap or lineshape. 
 These findings demonstrate that it is now possible to investigate the previously unexplored vibrational dynamics of host-guest systems that cannot be captured in small unit cells, leading to a rationalization of mode coupling and energy transfer pathways between host and guest vibrational modes. The ability to accurately predict vibrational properties in these systems opens new possibilities for engineering materials with tailored characteristics, such as optimized vibrational lifetimes or energy transfer pathways.
 
 \rev{While our results show promising extrapolation within PAH-based host–guest systems, extending this approach to chemically or structurally diverse host–guest systems will benefit from general ML potentials that can address a wide range of systems with reasonable accuracy and uncertainty-aware predictions that allow targeting refinement for accurate vibrational properties.}
 
 \section*{Data Availability Statement}
 \rev{The data and models used in this manuscript are available in Ref.~\cite{gurlek_2025_17100564}.}

\section*{Acknowledgements}
We acknowledge support from the Cluster of Excellence `CUI: Advanced Imaging of Matter'- EXC 2056 - project ID 390715994, BiGmax, the Max Planck Society’s Research Network on Big-Data-Driven Materials-Science and the Max Planck-New York City Center for Non-Equilibrium Quantum Phenomena. The Flatiron Institute is a division of the Simons Foundation. We also acknowledge support from the European Union’s Horizon Europe research and innovation programme under the Marie Skłodowska-Curie Doctoral Networks grant agreement No. 101118915 – TIMES. S.S. and P.L. acknowledge support from the UFAST International Max Planck Research School.

\section*{Author Contributions}
 B.G. and S.S. trained the potentials, implemented workflows, calculated, plotted and analysed results. B.G., S.S., P.L. and M.R. discussed and interpreted the results. B.G. and M.R. designed and supervised research. M.R. and A.R. acquired funding. B.G. and S.S. wrote the first draft of the manuscript. B.G. and M.R. finalized writing the manuscript with input from all authors.

\section*{Competing Interests}
The authors declare no competing interests.

\section*{Methods}\label{sec:Methods}

\subsection*{Computational Details}\label{App:comp_details}
The reference calculations for geometry relaxations, atomic forces, energies, and stresses of the molecular crystal structures are performed at the DFT level. All DFT calculations are conducted using the FHI-aims~\cite{Blum2009} and VASP~\cite{kresse1993ab,kresse1994ab,kresse1996efficiency,kresse1996efficient} codes, employing the Perdew-Burke-Ernzerhof (PBE) exchange-correlation functional~\cite{Perdew1996} with \rev{van der Waals corrections (vdW)~\cite{Tkatchenko2012, Tkatchenko2009} for FHI-AIMS and VASP, respectively}. We confirm that the relaxation algorithms in both VASP and FHI-aims produce vibrational frequency calculations that differ by no more than $\wavenumber{1.9}$, $\wavenumber{1.3}$, and $\wavenumber{4.6}$ for intermolecular, intramolecular, and C-H stretching modes, respectively. This ensures consistency and reliability for comparative analysis.

The on-the-fly machine learning algorithm in VASP dynamically constructs an accurate ML potential during AIMD simulations. It trains a kernel-based Gaussian process regression model using DFT-calculated energies, forces, and stresses for a selected subset of configurations. The algorithm assesses predictive uncertainty through Bayesian error estimates for each configuration encountered during the simulation~\cite{jinnouchi2019fly}. If the uncertainty exceeds a user-defined threshold, new DFT calculations are performed, and the resulting data are added to the training set to further refine the potential. \rev{The resulting DFT energies, forces and stresses are printed in the \texttt{ML\_AB} file, which contains the reference values for all selected configurations. These values serve as the ground-truth DFT data for training the potential, while the MLIP itself guides efficient sampling of relevant configurations. Structuring DFT reference data into the \texttt{ML\_AB} file allows VASP MLIP to be trained on external datasets without performing additional DFT calculations, as demonstrated in Methods.}

In the MACE machine learning architecture, atomic structures are represented as graphs, where nodes correspond to atoms in three-dimensional space, and any two nodes within a cutoff distance are connected. Multiple high body-order message-passing iterations are performed to update the features associated with each node in the network. The final atomic site energy prediction is obtained as a function of all node states generated throughout the iterations. In this work we used VASP version 6.4.2 and MACE version 0.3.10.

A $2\times 2\times 1$ Monkhorst-Pack k-mesh is used to perform single-point calculations. 

For VASP calculations, an energy cutoff of 1000~eV is applied to the plane-wave basis set, while a tight basis set is used for FHI-aims calculations. Geometry relaxations are performed within a fixed primitive cell, with a force convergence criterion of $4\times 10^{-4}$~eV/\AA. Phonon calculations are performed using the phonopy code with a $2 \times 2\times 2$ supercell~\cite{togo2023}. 

The key hyperparameters used to train the VASP MLIP, in addition to the default values, are $\text{ML\_MRB2}=12$, $\text{ML\_SION1}=0.3$ and $\text{ML\_WTSIF}=2$. Before the evaluations, the MLIP is refitted with sparsification  with $\text{ML\_MODE}=\text{refit}$. The active-learning step in Section~\ref{sec:VASPvsMACE} resulted in 1402 structures, and the learning iteration was terminated when negligible changes were observed in the number of structures, as well as in the mean forces and energies, as shown in Supplementary Fig.~21. 

The hyperparamteres used for training the MACE MLIPs are summarized in Table~\ref{tab:mace_hyperparameters}.
%%%%%%%%%%%%%%%%%%%%%%%%
\begin{table}[h!]
\centering
\caption{Hyperparameters for MACE MLIPs.}
\begin{tabular}{|l|c|}
\hline
\textbf{Hyperparameter}       & \textbf{Value} \\ \hline
Number of committee members          & 8               \\ \hline
Number of interactions        & 2              \\ \hline
Number of channels            & 256            \\ \hline
Maximum \( L \)               & 0              \\ \hline
Correlation                   & 3              \\ \hline
\( r_{\text{max}} \) (Å)      & 6.0            \\ \hline
Forces weight                 & 1000           \\ \hline
Energy weight                 & 10             \\ \hline
Train:Validation split         & 9:1            \\ \hline
\end{tabular}
\label{tab:mace_hyperparameters}
\end{table}
%%%%%%%%%%%%%%%%%%%%%%%%%%%%%%
%%%%%%%%%%%%%%%%%%%%%%%%
\begin{table}[h!]
\centering
\caption{Number of acene molecular crystals in generalized potentials}
\begin{tabular}{|l|c|c|c|c|}
\hline
\textbf{MLIP}       & \textbf{Naph.} &  \textbf{Anth.} & \textbf{Tetra.}& \textbf{Penta.}\\ \hline
N-MLIP & 450 & - & - & - \\ \hline
G-MLIP 1 & 450 & 125 & - & - \\ \hline
G-MLIP 2& 450 & 125 & 125 & - \\ \hline
G-MLIP 3& 450 & 125 & 125 & 125 \\ \hline
\end{tabular}
\label{tab:acenes_generalised_MLIP}
\end{table}
%%%%%%%%%%%%%%%%%%%%%%%%%%%%%%
The primitive lattice vectors and space groups of the molecular crystals used for training the generalized MACE MLIPs are provided in Tables~\ref{tab:naph_lattice}, ~\ref{tab:anth_lattice}, ~\ref{tab:tetra_lattice}, ~\ref{tab:penta_lattice}.
The dataset pools used for training generalized MACE MLIPs for acenes were generated as follows: For naphthalene, the dataset pool was created by performing $50$ ps ab-initio MD simulations on a $1 \times 2 \times 2$ supercell using FHI-aims at temperatures of $80$ K, $120$ K, $150$ K, $220$ K, and $295$ K ~\cite{capelli2006molecular}. 
For anthracene, the dataset pool was generated for a $1 \times 2 \times 2$ supercell using the universal forcefield MACE-OFF~\cite{kovacs2023maceoff} at temperatures of $100$ K and $295$ K. For tetracene and pentacene, the dataset pools were generated for $1 \times 1 \times 1$ cells using MACE-OFF at temperature $295$ K. The total number of structures in the dataset pool for each molecular crystal is listed in Table~\ref{tab:dataset_pool}. Single-point calculations for the selected structures in active learning iterations were performed using FHI-aims. 

\rev{For both MACE and VASP MLIPs, vdW interactions are implicitly accounted for in the results presented in the main text, whereas in Supplementary Note~3, we include them explicitly, similar to other models~\cite{Deringer2020,Anstine2025}.}  

%%%%%%%%%%%%%%%%%%%%%%%%
\begin{table}[h!]
\centering
\caption{Lattice parameters of the naphthalene crystal}
\begin{tabular}{|l|c|c|c|c|c|c|c|}
\hline
\textbf{\makecell{T (K)/ \\ CCDC entry}} & \textbf{a} & \textbf{b} & \textbf{c} & \textbf{$\alpha$} & \textbf{$\beta$} & \textbf{$\gamma$} & \textbf{Space group} \\ \hline

\makecell{80~\cite{capelli2006molecular} \\ NAPHTA33} 
& \makecell{8.10 \\ ~} & \makecell{5.94 \\ ~} & \makecell{8.64 \\ ~} 
& \makecell{90 \\ ~} & \makecell{124.46 \\ ~} & \makecell{90 \\ ~} 
& \makecell{P $2_{1}$/a (14) \\ ~} \\ \hline

\makecell{120~\cite{capelli2006molecular} \\ NAPHTA27} 
& \makecell{8.12 \\ ~} & \makecell{5.94 \\ ~} & \makecell{8.65 \\ ~} 
& \makecell{90 \\ ~} & \makecell{124.34 \\ ~} & \makecell{90 \\ ~} 
& \makecell{P $2_{1}$/a (14) \\ ~} \\ \hline

\makecell{150~\cite{capelli2006molecular} \\ NAPHTA34} 
& \makecell{8.14 \\ ~} & \makecell{5.95 \\ ~} & \makecell{8.65 \\ ~} 
& \makecell{90 \\ ~} & \makecell{124.08 \\ ~} & \makecell{90 \\ ~} 
& \makecell{P $2_{1}$/a (14) \\ ~} \\ \hline

\makecell{220~\cite{capelli2006molecular} \\ NAPHTA35} 
& \makecell{8.19 \\ ~} & \makecell{5.96 \\ ~} & \makecell{8.66 \\ ~} 
& \makecell{90 \\ ~} & \makecell{123.57 \\ ~} & \makecell{90 \\ ~} 
& \makecell{P $2_{1}$/a (14) \\ ~} \\ \hline

\makecell{295~\cite{capelli2006molecular} \\ NAPHTA36} 
& \makecell{8.25 \\ ~} & \makecell{5.98 \\ ~} & \makecell{8.67 \\ ~} 
& \makecell{90 \\ ~} & \makecell{122.72 \\ ~} & \makecell{90 \\ ~} 
& \makecell{P $2_{1}$/a (14) \\ ~} \\ \hline

\end{tabular}
\label{tab:naph_lattice}
\end{table}

%%%%%%%%%%%%%%%%%%%%%%%%%%%%%%
\begin{table}[h!]
\centering
\caption{Lattice parameters of the anthracene crystal}
\begin{tabular}{|l|c|c|c|c|c|c|c|}
\hline
\textbf{\makecell{T (K)/ \\ CCDC entry}} & \textbf{a} & \textbf{b} & \textbf{c} & \textbf{$\alpha$} & \textbf{$\beta$} & \textbf{$\gamma$} & \textbf{\makecell{Space \\ group}}  \\ \hline

\makecell{100~\cite{asher2020anharmonic} \\ ANTCEN25} 
& \makecell{9.28 \\ ~} & \makecell{5.99 \\ ~} & \makecell{8.41 \\ ~} 
& \makecell{90 \\ ~} & \makecell{102.52 \\ ~} & \makecell{90 \\ ~} 
& \makecell{P $2_{1}$/c (14) \\ ~} \\ \hline

\makecell{293~\cite{asher2020anharmonic} \\ ANTCEN26} 
& \makecell{9.45 \\ ~} & \makecell{6.00 \\ ~} & \makecell{8.54 \\ ~} 
& \makecell{90 \\ ~} & \makecell{103.51 \\ ~} & \makecell{90 \\ ~} 
& \makecell{P $2_{1}$/c (14) \\ ~} \\ \hline

\end{tabular}
\label{tab:anth_lattice}
\end{table}

%%%%%%%%%%%%%%%%%%%%%%%%%%%%%
\begin{table}[h!]
\centering
\caption{Lattice parameters of the tetracene crystal}
\begin{tabular}{|l|c|c|c|c|c|c|c|}
\hline
\textbf{\makecell{T (K)/ \\ CCDC entry}} & \textbf{a} & \textbf{b} & \textbf{c} & \textbf{$\alpha$} & \textbf{$\beta$} & \textbf{$\gamma$} & \textbf{\makecell{Space \\ group}} \\ \hline

\makecell{295~\cite{campbell1962crystal} \\ TETCEN} 
& \makecell{7.90 \\ ~} & \makecell{6.03 \\ ~} & \makecell{13.53 \\ ~} 
& \makecell{100.3 \\ ~} & \makecell{113.2 \\ ~} & \makecell{86.3 \\ ~} 
& \makecell{P $\overline{1}$ (2) \\ ~} \\ \hline

\end{tabular}
\label{tab:tetra_lattice}
\end{table}

%%%%%%%%%%%%%%%%%%%%%%%%%%%%%%
\begin{table}[h!]
\centering
\caption{Lattice parameters of the pentacene crystal}
\begin{tabular}{|l|c|c|c|c|c|c|c|}
\hline
\textbf{\makecell{T (K)/ \\ CCDC entry}} & \textbf{a} & \textbf{b} & \textbf{c} & \textbf{$\alpha$} & \textbf{$\beta$} & \textbf{$\gamma$} & \textbf{\makecell{Space \\ group}}  \\ \hline

\makecell{295~\cite{campbell1962crystal} \\ PENCEN} 
& \makecell{7.90 \\ ~} & \makecell{6.06 \\ ~} & \makecell{16.01 \\ ~} 
& \makecell{101.9 \\ ~} & \makecell{112.6 \\ ~} & \makecell{85.8 \\ ~} 
& \makecell{P $\overline{1}$ (2) \\ ~} \\ \hline

\end{tabular}
\label{tab:penta_lattice}
\end{table}

%%%%%%%%%%%%%%%%%%%%%%%%
\begin{table}[h!]
\centering
\caption{Dataset pool for molecular crystals. Only a subset is used for training the potentials.}
\begin{tabular}{|l|c|}
\hline
\textbf{Molecular crystal}       & \textbf{Number of structures} \\ \hline
Naphthalene          & 36859               \\ \hline
Anthracene        & 200000              \\ \hline
Tetracene            & 183123            \\ \hline
Pentacene               & 85578            \\ \hline
\end{tabular}
\label{tab:dataset_pool}
\end{table}
%%%%%%%%%%%%%%%%%%%%%%%%%%%%%%
%%%%%%%%%%%%%%%%%%%%%%%%%%%%%%

\subsection*{Committee-based active learning strategy}\label{App:committee}

To enhance training data efficiency and improve MLIP accuracy, we implemented a committee-based active learning strategy~\cite{schran2020committee, sivaraman2020machine, stolte2024random, kulichenko2023uncertainty, jinnouchi2020fly}. A committee of $8$ MACE MLIPs was trained simultaneously using the same dataset. The only differences among them were the initialization of weight parameters and the random split between the training and validation sets. All other hyperparameters were kept same throughout the active learning process and are detailed in Table~\ref{tab:mace_hyperparameters}. This approach enables each MLIP to capture different landscapes of the potential energy surface, enhancing the diversity in predictions.

A pool of atomic structures for acenes was generated at different temperatures, a detailed description is provided in Computational Details. The workflow for the committee-based active learning begins with selecting a small subset of labeled data from the pool to train the ensemble of MLIPs. Each MLIP is trained until the training and validation errors stabilize at sufficiently low values for both energy and forces. Following this training phase, long MD simulations are performed using the mean of the committee forces to assess the stability of the MLIPs across various temperature ranges and supercell sizes. For all MACE MLIPs, we use the accuracy of the $\Gamma$-point phonon frequencies, computed using the mean of the committee forces, as the stopping criterion for active learning, ensuring consistency with the reference DFT method. The number of AL-iterations, the corresponding number of training structures, and the achieved errors in energies and forces for the G-MLIPs are reported in Supplementary Tables~1-5. The total number of acene molecular structures in generalized potentials is also summarized in Table~\ref{tab:acenes_generalised_MLIP}.

Supplementary Figure~22 presents the maximum and mean errors in $\Gamma$-point phonon frequencies observed throughout the active-learning steps of G-MLIPs. Additionally, for comparison with the stopping criterion employed in VASP MLIP, we evaluate the mean and maximum energy and force errors per active-learning step for N-MLIP, as shown in Supplementary Figure~22. Once the initial training is complete, the committee of MLIPs makes predictions on the dataset pool. The 25 most uncertain atomic structures, based on the standard deviation of the predicted energies, are then selected. Their energies and forces are recomputed using the reference method and subsequently added to the training set. The committee is retrained on this expanded dataset, and the process is repeated to iteratively improve the model's performance and accuracy.

\subsection*{Uncertainty estimation and propagation to harmonic phonons}\label{App:uncertainity_estimation}

In committee of $M$ machine learning potentials, for any molecular structure $A$, the mean of committee is given by $\overline{y}(A)$ and the committee uncertainty is represented by its standard deviation $\sigma(A)$ ~\cite{imbalzano2021uncertainty, schran2020committee}. The active learning strategy in this work uses the committee uncertainty on energy to select new training structure from the pool.

For a given atomic structure $A$, we calculated the mean committee prediction, $\overline{y}(A)$, and the associated committee uncertainty, $\sigma(A)$, along with the reference values $y_\text{ref}(A)$ for $M$ committee members. Due to the limited number of training structures available to each committee member, the conditional probability distribution $P(y_\text{ref}(A) \lvert A)$ may deviate from an ideal Gaussian distribution. To account for this, the uncertainty is rescaled using a factor $\alpha$, computed as~\cite{imbalzano2021uncertainty}
\begin{align}
    \alpha^{2} \equiv -\frac{1}{M} +\frac{M-3}{M-1} \frac{1}{N_{test}} \sum_{A \in test} \frac {\lvert y_{\text{ref}}(A)-\overline{y}(A)\rvert^{2}}{\sigma^{2}(A)},\label{eq:scaling_alpha}
\end{align}
where $N_{test}$ represents the total number of structure in the test set, and the expression incorporates corrections for biases arising from the finite number of committee members. Using this scaling factor, we rescale the predictions of each committee member as 
\begin{align}
    y_i^{\prime}(A)=\overline{y}(A) + \alpha[y_i(A) - \overline{y}(A)],
    \label{eq:distribution}
\end{align}
which are used to compute vibrational properties, enabling robust uncertainty estimation due to the finite sampling of atomic configuration space.

Rescaling is performed on the committee forces for each member such that the uncertainty is accurately propagated onto the observables using equation Eq.~\ref{eq:distribution}
\begin{equation}
    \mathbf{F}_i^\prime(A) = \mathbf{\overline{F}}(A) + \alpha \lbrack \mathbf{F}_i(A) - \mathbf{\overline{F}}(A) \rbrack, \label{eq:scaled-force}
\end{equation}

We determine $\alpha$ for forces as $2.8$, on a validation set of 2100 diverse $1\times 2\times 2$ naphthalene crystal structures selected using FPS ~\cite{cersonsky2021improving, fps_eldar}.

To propagate this uncertainty to the harmonic phonons, we compute the phonon frequencies individually for each committee member using these scaled forces. This method enables the intrinsic propagation of uncertainty from the forces to the squared-phonon frequencies as
\begin{equation}
     \mathbf{F}^{\prime}_i(A) \longrightarrow \left[\omega_i(A)\right]^2.
\end{equation}

The phonon frequency prediction of committee, and the committee uncertainty are given as
\begin{equation}
    \overline{\omega^{\prime}} = \frac{1}{M} \sum^{M}_{i=1} \omega_i (A), \label{eq:omega-comm}
\end{equation}

\begin{equation}
    \sigma(\omega^{2}) = \sqrt{\frac{1}{M-1} {\sum^{M}_{i=1} \left[ \omega_i^2 - \overline{\omega^{\prime}} ^{2}\right]^2}}. \label{eq:sigmaomega-comm}
\end{equation}
Since forces are directly proportional to the square of the frequencies, the uncertainty in forces propagates to the square of the frequencies. $\mathbf{F} \propto \omega^2 \rightarrow d\mathbf{F}/d\omega = 2  \omega$, so $\sigma_{\omega} = \sigma_{\omega^2}/|2\omega|$.

\subsection*{Uncertainty propagation to vibrational density of states}\label{App:uncertainity_vdos}

The VDOS is calculated for each committee member $i$ with the corresponding scaled forces $\mathbf{F}_i$ from the velocity auto-correlation function (VACF)  
\begin{equation}
    C_{\mathbf{F}_i}(t)  = \lim_{T \to \infty} \frac{1}{T - t} \sum_j\int_{0}^{T - t} m_j\mathbf{v}^j_{\mathbf{F}_i}(\tau) \cdot \mathbf{v}^j_{\mathbf{F}_i}(\tau + t) \, d\tau, \label{eq:vdos}
\end{equation}
where $\mathbf{v}^j_{\mathbf{F}_i}(t)$ is a shorthand notation for the velocity of the $j^\text{th}$ atom in Cartesian coordinates at time $t$, as predicted by the $i^\text{th}$ committee member. The VDOS spectrum ($C_{\mathbf{F}_i}(\omega)$) is obtained from the Fourier transform of the corresponding VACF. Atomic trajectories are obtained from NVE simulations following equilibration runs in the NVT ensemble. For the naphthalene molecular crystal, 100 simulations per committee member are run for 20 ps each at 80 K. In  addition, with mean of the committee forces, 100 simulations  are run for 20 ps at 80 K for naphtalene molecular crystal, while for the host-guest system, 27 simulations are conducted for 15 ps each at 100 K.

The statistical error, $\sigma_{\mathrm{stat}}$ is calculated by block averaging of VDOS over NVE runs, i.e., 
\begin{equation}
\sigma_{\text{stat}}(\omega) = \sqrt{\frac{1}{N-1}\sum_{k=1}^N \left[\overline{C_{\mathbf{\overline F}}(\omega)} - C_{\mathbf{\overline F}}^k(\omega) \right]^{2}}, \label{eq:sigmastat}
\end{equation}
where $\overline{C_{\mathbf{\overline F}}(\omega)}$ is the block averaged VDOS calculated with mean of the committee forces ${\mathbf{\overline F}}$ and  $C_{\mathbf{\overline F}}^k(\omega)$ is the VDOS corresponding to $k^\text{th}$ trajectory obtained with the mean of the committee forces ${\mathbf{\overline F}}$, and $N$ is the total number of NVE runs.

We estimate the total error in VDOS by calculating the standard deviation across all committee members as
\begin{equation}
    \sigma_{\mathrm{total}}(\omega) = \sqrt{\frac{1}{(M-1)(N-1)} \sum^{M}_{k=1}\sum^{M}_{i=1} \left[ C_{\mathbf{F}_i}^{ik}(\omega) - \overline{C_{\mathbf{F}_i}(\omega)}\right]^{2}},\label{eq:total_error}
\end{equation}
where, $N = 100$, and $M=8$. 

Committee uncertainty and the corresponding committee error are computed as,
\begin{equation}
    \sigma_{\mathrm{com}}(\omega) = \sqrt{\left[\sigma_{\mathrm{total}}(\omega)\right]^2 - \left[\sigma_{\mathrm{stat}}(\omega)\right]^2},
\label{eq:com_uncer}
\end{equation}
\begin{equation}
\sigma^{\mathrm{e}(\omega)}_{\mathrm{com}} = \frac{\sigma_{\mathrm{com}}(\omega)}{\sqrt{M}}.
\label{eq:com_error}
\end{equation}

\subsection*{Normal-Mode Projected VDOS}\label{App:ProjectedVDOS}
We follow the procedure outlined in Ref.~\cite{Sun-2014} to compute the normal-mode projected VV-ACF and the corresponding power spectra. We first calculate the normal-mode projected atomic velocities ($v_{s}(t)$) from the AIMD trajectories as
\begin{align}
     v_{s}(t) = \sum_j \sqrt{m_j} \mathbf{e}_j^*(s)\cdot \mathbf{v}(t),\label{eq:projected_velocities}
\end{align}
where $m_j$ is the mass of $j^\text{th}$ atom, $\mathbf{e}_j^*(s)$ is the polarization vector of the harmonic phonon at $\Gamma$-point, and $\mathbf{v}(t)$ are the atomic velocities in $x, y$ and $z$ directions. The normal-mode projected VDOS can then be calculated via Eq.~\eqref{eq:vdos} as $C_s(\omega)=\mathcal{F}\{\langle v_{s}(0) v_{s}(t)\rangle\}$. The total VDOS can be obtained via $C(\omega)=\sum_s C_s(\omega)$.

Here, we calculate the contributions of host and guest atoms to the VDOS by separating the normal-mode projected velocities as $v_{s}(t)=v_{s}^\text{h}(t)+v_{s}^\text{g}(t)$, where the last two terms are the contributions of host and guest atoms into the projected velocities, i.e., $v_{s}^\text{host(guest)}(t) = \sum_{j\in \text{host(guest)}} \sqrt{m_j} \mathbf{e}_j^*(s)\cdot \mathbf{v}(t)$. Hence the power spectrum can be decomposed to three terms as 
\begin{align}
    C_s(\omega)=C_s^\text{host}(\omega)+C_s^\text{guest}(\omega)+C_s^\text{cross}(\omega),
\end{align}
 where the host(guest)-projected VDOS is $C_s^\text{host(guest)}(\omega)=\mathcal{F}\{\langle v_{s}^\text{host(guest)}(0) v_{s}^\text{host(guest)}(t)\rangle\}$, and the cross-correlated VDOS is $C_s^\text{cross}(\omega)=\mathcal{F}\{\langle v_{s}^\text{guest}(0) v_{s}^\text{host}(t)\rangle+\langle v_{s}^\text{host}(0) v_{s}^\text{guest}(t)\rangle\}$. The cross terms can be interpreted as a measure of the coupling between host and guest normal modes in the harmonic phonon basis.  

%
% ****** End of file apssamp.tex ******


\clearpage
% \documentclass[aps, physrev, groupedaddress,superscriptaddress, amsmath,amssymb,floatfix]{revtex4-2}
% \usepackage{graphicx}
% \usepackage{dcolumn}
% \usepackage{bm}
% \usepackage{hyperref}
% \usepackage{braket}
% \usepackage{color}
% \usepackage{comment}
% \usepackage{sidecap}
% \usepackage{multirow}
% \setlength {\marginparwidth }{2cm}
% \newcommand{\wavenumber}[1]{#1~\text{cm}^{-1}}
% \newcommand{\rev}[1]{{\color{black} #1}}
% \usepackage{todonotes}
% \usepackage{bm}
% %\newcommand{\todoo}[1]{{\color{blue}\bf  #1}}
% %\newcommand{\BG}[1]{{\color{red}\bf  #1}}
% %\newcommand{\wavenumber}[1]{#1~\text{cm}^{-1}}
% \begin{document}

\clearpage
\onecolumngrid
\begin{center}
    {\Large \bfseries Supplementary Information: Accurate Machine Learning Interatomic Potentials for Polyacene Molecular Crystals: Application to Single Molecule Host-Guest System \par}
    \vspace{0.8cm}

    Burak Gurlek\textsuperscript{1,*},  
    Shubham Sharma\textsuperscript{1,*},  
    Paolo Lazzaroni\textsuperscript{1},  
    Angel Rubio\textsuperscript{1,2},  
    Mariana Rossi\textsuperscript{1,$\dagger$}  
    \vspace{0.5cm}

    {\small
    \textsuperscript{1} Max Planck Institute for the Structure and Dynamics of Matter and Center for Free-Electron Laser Science,  
    Luruper Chaussee 149, 22761 Hamburg, Germany \\[0.2cm]
    \textsuperscript{2} Initiative for Computational Catalysis (ICC), The Flatiron Institute,  
    162 Fifth Avenue, New York, New York 10010, USA
    } \\[0.5cm]

    {\small *These two authors contributed equally} \\
    {\small $\dagger$ mariana.rossi@mpsd.mpg.de}
\end{center}

\vspace{1cm}

\setcounter{section}{0}
\setcounter{figure}{0}
\setcounter{table}{0}
\renewcommand{\thesubsection}{Supplementary Note~\arabic{subsection}} % Use regular numbering for sections
\renewcommand{\theequation}{Supplementary Equation~\arabic{equation}}
\captionsetup[figure]{justification=justified,labelfont=bf, labelformat=simple, labelsep=colon, name={Supplementary Figure}}
\captionsetup[table]{justification=justified,labelfont=bf, labelformat=simple, labelsep=colon, name={Supplementary Table}}
\renewcommand{\thetable}{\arabic{table}}

\subsection{Comparing the Performance of MACE and VASP MLIPs on VASP Tranining Dataset}\label{sec:vaspvsmace}
The performance of the VASP nad MACE MLIPs is best depicted by the correlation plots shown in Supplementary Figure s~\ref{fig:vaspvsmace}(a, b), obtained by predicting energies and forces on the test set (see Methods). Both MLIPs show strong performance in predicting atomic force components. Regarding energies, the MACE MLIP captures the correlation slope but overestimates most energies, whereas the VASP MLIP achieves accurate energy predictions.
We observed that the MACE MLIP accurately predicts energies primarily for structures at temperatures close to its training temperature of $295$~K. This limitation appears related to the MACE model’s difficulty in capturing stress components (version 0.3.10). To verify this argument, we trained a new VASP MLIP on the same dataset without including stress fitting. The predictions of this new model on the same test set shows a very a similar energy overestimation trend~(see Supplementary Figure ~\ref{fig:SI_VASP_NoStress}). We note, however, that this behaviour of MACE was not changed by increasing the loss-function weight on stresses and the exact source of this shortcoming remains unexplained.  
%%%%%%%%%%%%%%%%%%%%%%%%%%%%%
\begin{figure}[ht]
\centering
    \includegraphics[width=16.5cm]{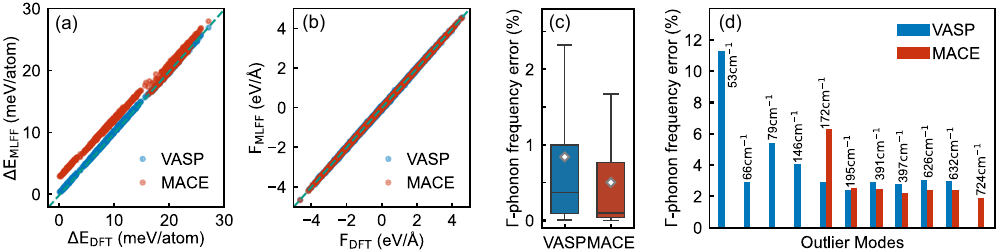}
        \caption{\textbf{Comparison of VASP and MACE single potential MLIPs for the naphthalene molecular crystal.}~\textbf{a} Correlation plot of relative energies, $\Delta E_\text{DFT(MLFF)}=E_\text{DFT(MLFF)}-E_\text{DFT}^\text{min}$, and \textbf{(b)} forces predicted by DFT and MLIPs.~\textbf{c} Error box plot for harmonic phonon frequencies ($\Gamma$-point) obtained with MLIPs.~\textbf{d} Outliers identified from the box plot in \textbf{c}. The wavenumber labels refer to the associated modes in the reference DFT calculations using VASP. The training isbased on a single-temperature at $295$K.}
        \label{fig:vaspvsmace}
\end{figure}

To get insight into the performance of VASP and MACE MLIPs for vibrational properties, we next compare the harmonic $\Gamma$-point phonon frequencies of a naphthalene molecular crystal as calculated with the MLIP potentials and with DFT. In Supplementary Figure ~\ref{fig:vaspvsmace}(c), we plot the corresponding percentage errors of all frequencies relative to the reference DFT calculations. These errors show a skewed distribution for both VASP and MACE MLIPs with the mean percentage (absolute) errors of $0.84\%$ ($\wavenumber{4.34}$) and $0.50\%$ ($\wavenumber{2.56}$) and with non-outlier maximums of $2.32\%$ ($\wavenumber{11.54}$) and $1.67\%$ ($\wavenumber{4.95}$), respectively. This indicates better performance by the MACE MLIP in predicting vibrational properties, consistent with the training errors presented in Table~\ref{tab:errors_vaspmace}. We note that not all normal modes calculated with the MLIPs follow the same ordering in terms of ascending frequency values as they do in DFT. Reordering MLIP-predicted frequencies based on the cosine similarity between MLIP and DFT phonon eigenvectors leads to even larger errors~(see Supplementary Figure ~\ref{fig:SI_VASPvsMACE_Angle}).
 %%%%%%%%%%%%%%%%%%%%%%%%%%%%%%%%%5

\begin{figure*}[!h]
\centering
\includegraphics[width=8cm]{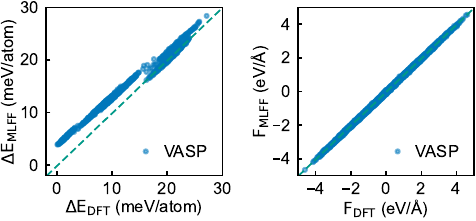}
\caption{\textbf{Comparing the performance of VASP MLIP without stress fitting.}~\textbf{a} Relative energy, $\Delta E_\text{DFT(MLFF)}=E_\text{DFT(MLFF)}-E_\text{DFT}^\text{min}$, and force \textbf{(b)} predictions of the VASP MLIP trained on the same training set and evaluated on the same test set as in Supplementary Figure ~\ref{fig:vaspvsmace}.}
\label{fig:SI_VASP_NoStress}
\end{figure*}
%%%%%%%%%%%%%%%%%%%%%%%%%%%%%%%%%%
We also present outlier vibrational modes in terms of percentage errors in Supplementary Figure ~\ref{fig:vaspvsmace}(d), where the VASP MLIP exhibits the largest errors for intermolecular vibrational modes, with frequencies below $\wavenumber{145-150}$. Conversely, the largest percentage errors for the MACE MLIP occur in the first few intramolecular vibrational modes. A more detailed examination of absolute errors for individual phonon modes reveals that the MACE MLIP performs better for intermolecular vibrational modes, with mean errors of $\wavenumber{2.48}$ for VASP and $\wavenumber{0.77}$ for MACE MLIPs. 
Moreover, the performance of the VASP MLIP declines for intramolecular vibrational modes, with a mean frequency error of $\wavenumber{4.48}$, while the MACE MLIP shows a similar trend, with a mean frequency error of $\wavenumber{2.71}$. Notably, for intramolecular vibrational modes with frequencies in between $\wavenumber{500-1000}$, corresponding to C-H bending and ring deformation modes, both MLIPs exhibit similar errors~[see Supplementary Figure ~\ref{fig:SI_VASPvsMACE_Phonon}(a)], suggesting challenges in accurately capturing forces related to bending and torsional forces. However, for the C-H stretch modes around $\wavenumber{3100}$, both MLIPs perform better than their overall averages, with mean errors of $\wavenumber{1.58}$ and $\wavenumber{0.89}$, respectively.
%%%%%%%%5
\subsection{Tranining VASP MLIP with the MACE MLIP-committee 
dataset}\label{section:SI_VASP_withMACE}
To test the data transferablity of the VASP MLIP, we trained a VASP MLIP on the same dataset used to produce the MACE MLIP committee, keeping the same parameters used in building previous VASP MLIPs. Supplementary Table~\ref{tab:errors_vasp_macedataset} presents the training and validation errors on the testset (see Methods) and compares these with the errors from the MACE MLIP-multi relative to FHI-AIMS reference DFT calculations. The VASP MLIP exhibits slightly higher errors compared to previous VASP MLIPs (see Table~\ref{tab:errors_vaspmace}~and~\ref{tab:errors_bestvaspmace}). Importantly, it does not perform as well as MACE MLIP-multi despite using the same atomic configuration space. Moreover, the new VASP MLIP showed lower stability compared to previous model during a $2$ ns NVT-MD run. 

%%%%%%%%%%%%%%%%%%%%%%%%%%%%%
\begin{table}[h] % Table environment
    \centering % Center the table
    \caption{The training and validation errors of VASP MLIP, trained on MACE MLIP-committee training dataset. The results from MACE MLIP-committee is shown for reference. The errors are presented as RMSE for energies and forces.} 
    \label{tab:errors_vasp_macedataset} % Add label for referencing
    \begin{tabular}{ p{1.1cm} >{\centering\arraybackslash}p{1.6cm} >{\centering\arraybackslash}p{1.6cm} >{\centering\arraybackslash}p{1.6cm} >{\centering\arraybackslash}p{1.6cm} }
         & \multicolumn{2}{c}{Energy (meV/atom)} & \multicolumn{2}{c}{Force (meV/\AA)} \\ 
        \hline
         & Training & Test & Training & Test \\
         \hline
        VASP   & 0.1 &0.1    & 30.1 &26.4    \\ 
        MACE   & 0.08 &0.08  & 4.3 &4.4   \\ 
        \hline
    \end{tabular}
\end{table}
%%%%%%%%%%%%%%%%%%%%%%%%%%%%%%%%%%%%
In Supplementary Figure ~\ref{fig:vasp_macedataset}~(a), we further compare the $\Gamma$-phonon frequency errors of a naphthalene molecular crystal predicted by the VASP MLIP and MACE MLIP-multi, showing an order-of-magnitude larger mean error for VASP MLIP ($\wavenumber{2.13}$). The errors are even more pronounced for the outlier modes, especially for intermolecular modes, reaching up to $30\%$ as shown in Supplementary Figure ~\ref{fig:vasp_macedataset}~(b). Overall, these errors are also worse compared to those of VASP MLIP-multi trained on a similar dataset generated through VASP's active-learning algorithm. This results highlight the limitation of VASP MLIP algorithm in transferring datasets between different MLIPs for accurately predicting vibrational dynamics.
%%%%%%%%%%%%%%%%%%%%%%%%%%%%%%%%%%%
\begin{figure}[ht]
\centering
    \includegraphics[width=9cm]{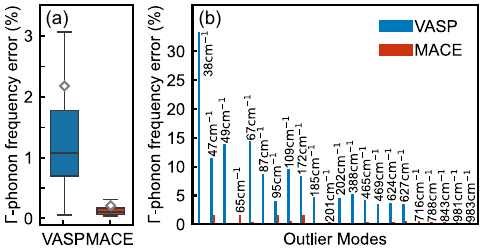}
        \caption{\textbf{Performance of VASP MLIP trained on the MACE MLIP-committee training data set for naphthalene molecular crystal}.~\textbf{a} Error box plot for $\Gamma$-phonons obtained with MLIPs.~\textbf{b} Outliers identified from the box plot in \textbf{a}. The wavenumbers of the modes correspond to those of the associated modes in reference DFT calculations. For reference, the MACE MLIP-committee results shown in Figure ~\ref{fig:bestvaspvsmace} replotted.}
        \label{fig:vasp_macedataset}
\end{figure}
%%%%%%%%%%%%%%%%%%%%%%%%%%%%%%%%%

%%%%%%%%%%%%%%%%%%%%%%%%%%%%
\begin{figure*}[!h]
\centering
\includegraphics[width=16.5cm]{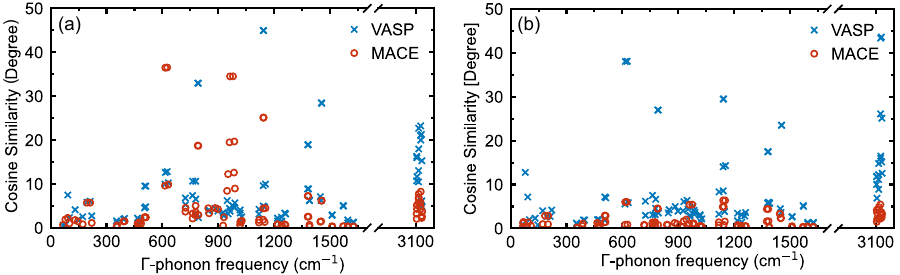}
\caption{\textbf{Cosine similarity measure between phonon eigenvectors predicted by VASP and MACE MLIPs.}~\textbf{a} VASP and MACE MLIP.~\textbf{b} VASP MLIP-multi and MACE MLIP-committee.}
\label{fig:SI_VASPvsMACE_Angle}
\end{figure*}

%%%%%%%%%%%%%%%%%%%%%%%5
\begin{figure*}[!h]
\centering
\includegraphics[width=16cm]{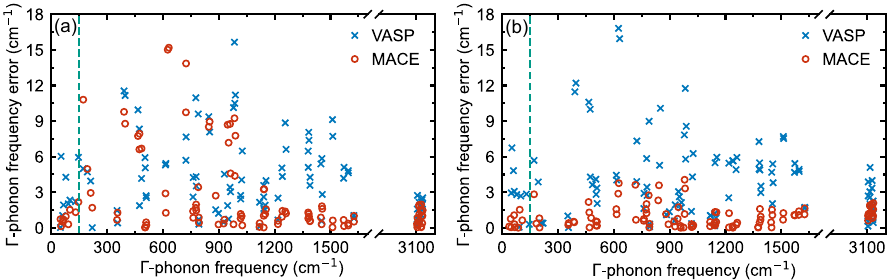}
\caption{\textbf{Absolute errors of $\Gamma$-phonon frequencies in naphthalene crystal as predicted by MLIPs and DFT.}~\textbf{a} Comparing VASP and MACE MLIP.~\textbf{b} Comparing VASP MLIP-multi and MACE MLIP-committee. The green line marks the boundary between intermolecular and intramolecular vibrational modes with frequency $\wavenumber{145}$.}
\label{fig:SI_VASPvsMACE_Phonon}
\end{figure*}
%%%%%%%%%%%%%%%%%%%%%%%%
% \begin{figure}[hb]
% \centering
%     \includegraphics[width=8 cm]{figures/Fig_SI_Phonon_FreqE_ComMeanvsRescaleF.pdf}
%         \caption{Comparison of mean phonon frequencies $\overline{\omega}^\prime$ (the mean prediction of the individual phonon frequency $\omega^{(i)^{\prime}}$ predictions from $i^\text{th}$ committee members, obtained by using their respective scaled forces $\vec{F}^{(i)^{\prime}}$) and $\overline{\omega}$, (the committee mean phonon frequencies computed using the mean forces $\vec{F}^{(i)^{\prime}}$ of the committee)  with respect to $\omega_{ref}$ for a $2$ $\times$ $2$ $\times$ $2$ supercell of naphthalene crystal.}
%         \label{fig:SI_Phonon_FreqE_ComMeanvsRescaleF}
% \end{figure}
%%%%%%%%%%%%%%%%%%%%%%%%
\begin{figure}[hb]
\centering
    \includegraphics[width=12 cm]{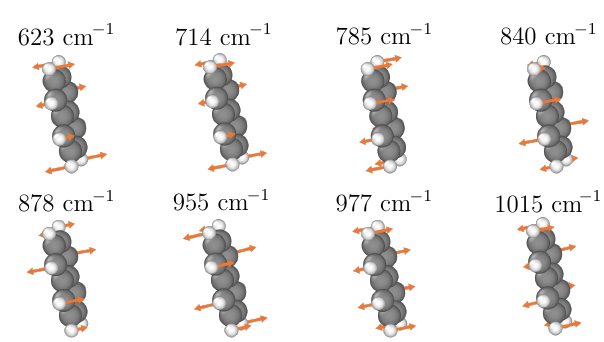}
        \caption{\textbf{Illustrations of the naphthalene phonon modes with large uncertainties in the predicted vibrational frequencies.} The modes with uncertainities greater than $\wavenumber{7}$ is plotted.}
        \label{fig:SI_Naphtalene_Phonons}
\end{figure}
%%%%%%%%%%%%%%%%%%%%%%5
\begin{figure}[hb]
\centering
    \includegraphics[width=10 cm]{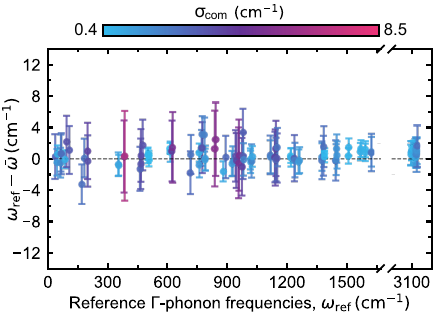}
        \caption{\textbf{Uncertainty estimations for the $\Gamma$-point phonon frequencies after an active-learning iteration on NMLIP.}~At $295$~K, 25 structures were generated by randomly combining normal mode displacements with frequencies in the range $\wavenumber{600} - \wavenumber{1000}$, weighted by their thermal amplitudes ($\sqrt{k_B T}/\omega$). Only modes with uncertainties larger than $\wavenumber{4}$ were considered (see Figure ~\ref{fig:PhononError_Propagation}(a)).}
        \label{fig:SI_NMLIP_ActiveLearning}
\end{figure}
%%%%%%%%%%%%%%%%%%%%%%5
\begin{figure}[hb]
\centering
    \includegraphics[width=17 cm]{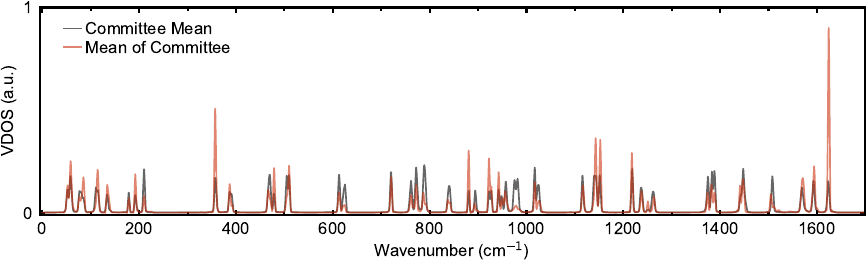}
        \caption{\textbf{Comparison of the mean of committee VDOS}. VDOS is computed from trajectories propagated with the mean of committee forces, and the committee mean VDOS, obtained by averaging VDOSs computed with the respective committee-scaled forces.}
        \label{fig:SI_CommMeanComm}
\end{figure}
%%%%%%%%%%%%%%%%%%%%%%5
\begin{figure}[hb]
\centering
    \includegraphics[width=17 cm]{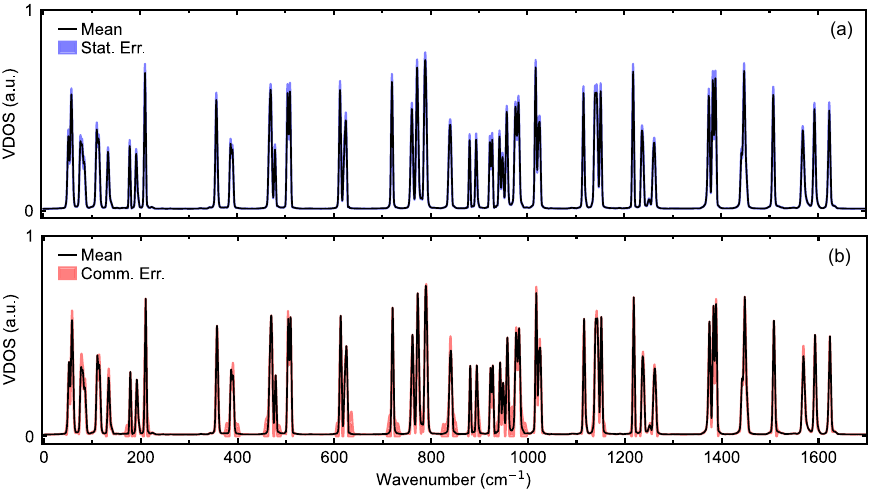}
        \caption{\textbf{Uncertainty propagation for the VDOS of the $1\times 1\times 1$ naphthalene molecular crystal.}~\textbf{a} Statistical and \textbf{b} committee errors, respectively.}
        \label{fig:SI_VDOS_Uncertainity}
\end{figure}

%%%%%%%%%%%%%%%%%%%%%%%%5
\begin{figure*}[!h]
\centering
\includegraphics[width=17cm]{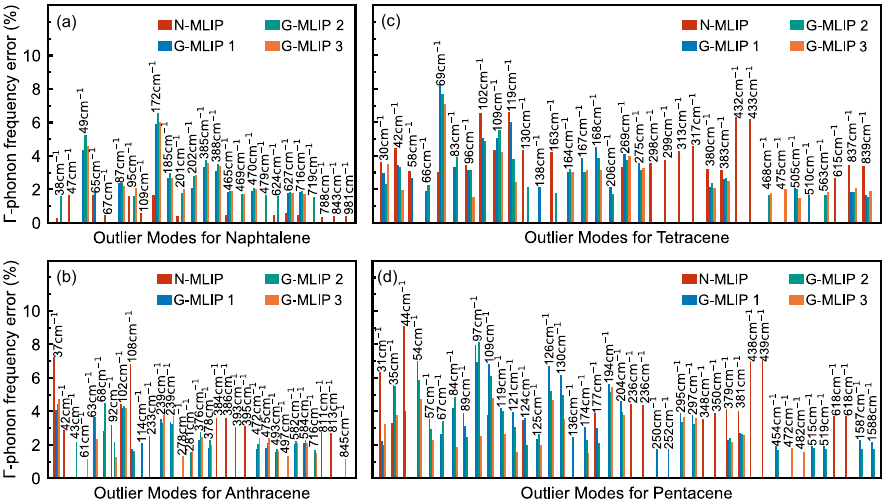}
\caption{\textbf{MLIP performance for $\Gamma$-point phonon frequencies across polyacene crystals}~Outlier phonon mode errors corresponding to the box plots in Figure ~\ref{fig:Fig_MultiMLFF} for the naphthalene \textbf{(a)}, anthracene \textbf{(b)}, tetracene \textbf{(c)}, and pentacene \textbf{(d)} molecular crystals.}
\label{fig:SI_MultiMLFF}
\end{figure*}
%%%%%%%%%%%%%%%%%%%%%%%%5
\begin{figure*}[!h]
\centering
\includegraphics[width=16cm]{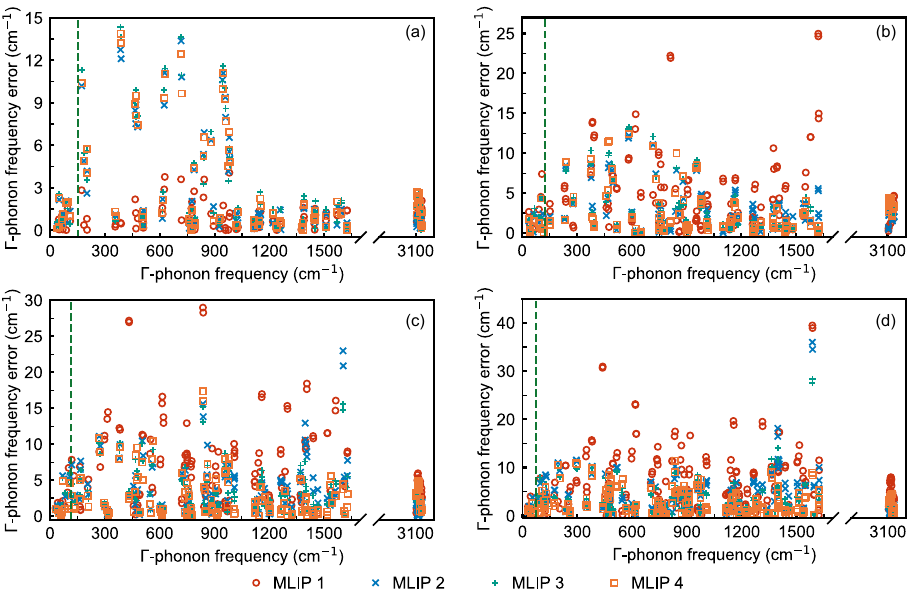}
\caption{\textbf{Absolute errors of phonon frequencies predicted by generalized MLIPs.}~\textbf{a} naphthalene, \textbf{(b)} anthracene, \textbf{c} tetracene and \textbf{d} pentacene molecular crystals. The green line marks the boundary between intermolecular and intramolecular vibrational modes with frequency $\wavenumber{145, 125, 115, 75}$, respectively.}
\label{fig:SI_Phonon_Generalized}
\end{figure*}
%%%%%%%%%%%%%%%%%%%%%%%%5
%%%%%%%%%%%%%%%%%%%%%%%%%%%%
\begin{figure*}[!h]
\centering
\includegraphics[width=0.7\textwidth]{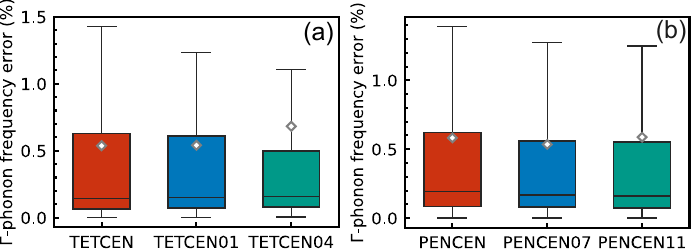}
\caption{\textbf{Percentage errors on harmonic $\Gamma$-point phonon frequencies calculated with the G-MLIP3 for different polymorphs.}~\textbf{a} tetracene and \textbf{(b)} pentacene molecular crystals. The potential was trained only with structures from the TETCEN (1269538) and PENCEN (1230799) polymorphs. The labeling of the different polymorphs follows the CCDC database entry and the database deposition numbers are given in parenthesis. The errors on TETCEN01 (114446), TETCEN04 (1844642), PENCEN07 (619979) and PENCEN11 (674030) are low and comparable to the errors obtained on the TENCEN and PENCEN polymorphs.}
\label{fig:SI_polymorphs}
\end{figure*}
%%%%%%%%%%%%%%%%%%%%%%%

\begin{figure*}[!h]
\centering
\includegraphics[width=17cm]{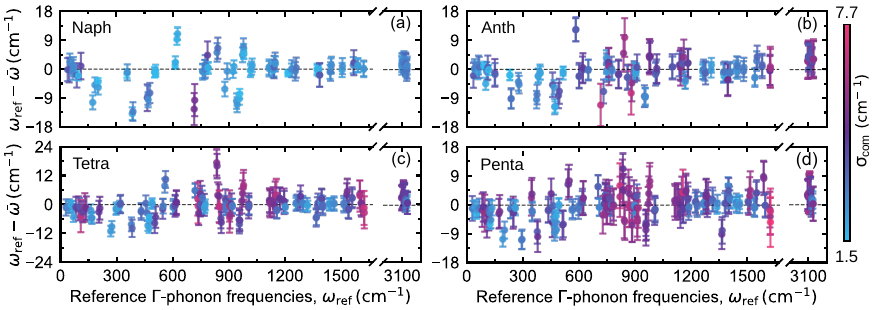}
\caption{\textbf{Uncertainty propagation for the $\Gamma$-point phonon frequencies predicted by G-MLIP$3$.}~\textbf{a} naphthalene, \textbf{(b)} anthracene, \textbf{c} tetracene and \textbf{d} molecular crystals. The color bar represents the standard deviation among committee members ($\sigma_\text{com}$). Interestingly, for the the uncertainties are more predictive of the real error, even if underestimations are still pronounced for a few modes. The region between $\wavenumber{600-900}$ shows the largest uncertainties across all crystals, hinting that G-MLIP3 struggles to capture benzene-ring deformations and CH-bend vibrations. As we showed in Supplementary Figure~\ref{fig:SI_NMLIP_ActiveLearning}, this can be solved by pruning the training set to better cover these motions, thus using the uncertainty quantification to further enhance the potential.}
\label{fig:SI_MultiFF_rev}
\end{figure*}
%%%%%%%%%%%%%%%%%%%%%%%%5

\begin{table}[h] % Table environment
    \centering % Center the table
    \caption{Absolute errors and statistical measures of $\Gamma$-phonon frequencies for naphthalene, anthracene, tetracene, and pentacene molecular crystals as predicted by generalized MLIPs. Units are in $\wavenumber{}$. ‘Inter.’ and ‘Intra.’ denote intermolecular and intramolecular, respectively.} 
    \label{tab:errors_multimlff} % Add label for referencing
    \begin{tabular}{ p{2.1cm} >{\centering\arraybackslash}p{2.6cm} >{\centering\arraybackslash}p{1.8cm} >{\centering\arraybackslash}p{1.8cm} >{\centering\arraybackslash}p{1.8cm} >{\centering\arraybackslash}p{1.8cm} }
        & &N-MLIP & G-MLIP 1 &G-MLIP 2&G-MLIP 3 \\ 
        \hline
         \multirow{8}{*}{Naphthalene} & Mean & 0.96 & 2.57 & 2.96 & 2.79\\ 
         & Maximum & 3.14 & 13.13 & 14.82 & 14.32 \\
         & Mean Inter. & 0.50 & 1.25 & 1.27 & 1.11\\
         & Maximum Inter. & 1.75 & 3.86 & 3.18 & 2.79\\
         & Mean Intra. & 0.72 & 2.67 & 3.11 & 2.93\\
         & Maximum Intra. & 3.14 & 13.13 & 14.82& 14.31\\
         & Mean C-H & 0.62 & 0.84 & 1.97 & 1.86\\
         & Maximum C-H & 1.41 & 2.16 & 3.45& 3.34\\
        \hline
         \multirow{8}{*}{Anthracene} & Mean & 4.82 & 2.84 & 2.86& 2.76\\ 
         & Maximum & 24.97 & 12.19 & 13.28 & 12.46 \\
         & Mean Inter. & 1.88 & 1.52 & 1.84 & 1.34\\
         & Maximum Inter. & 4.53 & 4.32 & 4.44 & 4.32\\
         & Mean Intra. & 4.98 & 2.93 & 2.94 & 2.86\\
         & Maximum Intra. & 24.97 & 12.19 & 13.28 &  12.46\\
         & Mean C-H & 2.53 & 2.29 & 3.17 & 3.38\\
         & Maximum C-H & 4.67 & 3.89 & 4.23 & 4.39\\
        \hline
        \multirow{8}{*}{Tetracene} & Mean & 6.50 & 3.52 & 3.17 & 2.70\\ 
         & Maximum & 40.08 & 22.97 & 15.54  & 17.32 \\
         & Mean Inter. & 1.31 & 2.01 & 1.82 & 1.26\\
         & Maximum Inter. & 2.10 & 5.61 & 5.30 & 4.90\\
         & Mean Intra. & 6.73 & 3.59 & 3.22 & 2.77\\
         & Maximum Intra. & 40.08 & 22.07 & 15.54  & 17.32\\
         & Mean C-H & 3.87 & 0.93 & 2.26 & 2.94\\
         & Maximum C-H & 5.95 & 1.89 & 4.40 & 4.85\\
        \hline
      \multirow{8}{*}{Pentacene} & Mean & 7.09 & 4.05 & 3.40 &  2.85\\ 
         & Maximum &  39.46 & 35.97 & 28.31 & 11.41 \\
         & Mean Inter. & 1.87 & 2.05 & 1.60 & 1.35\\
         & Maximum Inter. & 4.02 & 3.82 & 3.20 & 2.57\\
         & Mean Intra. & 9.42 & 4.11 & 3.45 & 2.90\\
         & Maximum Intra. & 39.46 & 35.97 & 28.31 & 11.41\\
         & Mean C-H & 5.34 & 0.88 & 1.74 & 2.85\\
         & Maximum C-H & 8.00 & 2.22 & 2.84 & 4.49\\
        \hline
    \end{tabular}
\end{table}

%%%%%%%%%%%%%%%%%%%%%%%%%%%%
\begin{figure*}[!h]
\centering
\includegraphics[width=16.5cm]{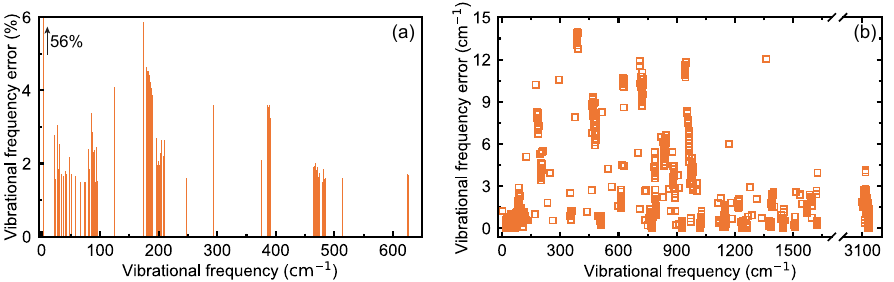}
\caption{\textbf{Errors in vibrational frequencies of the host-guest system predicted by G-MLIP3.}~\textbf{a} Percentage errors for outlier modes.~\textbf{b} Absolute errors across the entire spectrum.}
\label{fig:SI_hostguest_phononerror}
\end{figure*}
%%%%%%%%%%%%%%%%%%%%%%%
\begin{figure*}[!h]
\centering
\includegraphics[width=9cm]{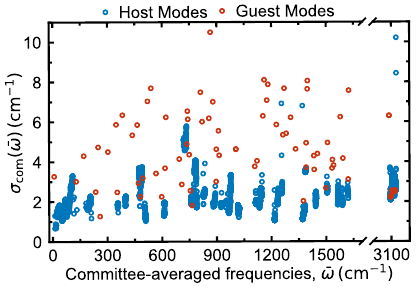}
\caption{\textbf{Uncertainity propagation for $Gamma$-point phonon frequencies of the host-guest system.}~The host-guest system is pentacene-doped $4 \times 4 \times 5$ naphthalene supercell. When projecting these uncertainties on host and guest, we find that modes with largest (outlier) uncertainty correspond to guest-dominated modes, which demonstrate comparatively less reliable predictions ($\bar\sigma_\text{com}(\bar\omega) \approx~\wavenumber{4.6}$) compared to host-dominated modes ($\bar\sigma_\text{com}(\bar\omega) \approx \wavenumber{2.2}$).}
\label{fig:SI_hostguest_uncertainity}
\end{figure*}
%%%%%%%%%%%%%%%%%%%%%%%
\clearpage
\subsection{Explicit van der Waals Energy correction for  NMLIP and G-MLIP-3}\label{section:SI_vdW_Correction}
\begin{figure*}[!h]
\centering
\includegraphics[width=15cm]{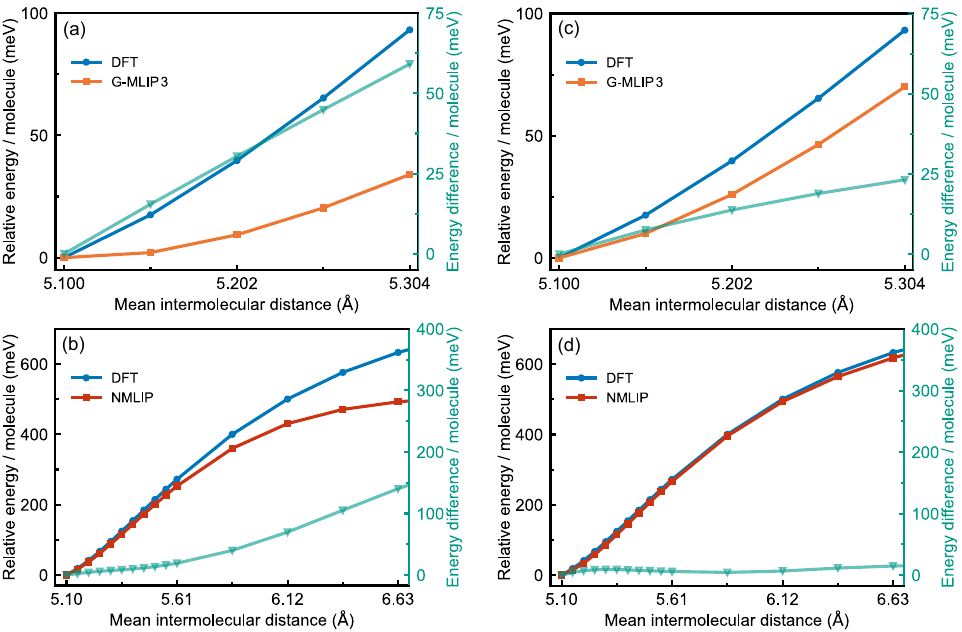}
\caption{\textbf{Benchmarking performance of MLIPs on long-range vdW interactions.}~Relative free energies predicted by DFT (blue circles), G-MLIP$3$ (orange squares) and NMLIP (red squares) as intermolecular distances are expanded by uniformly increasing the spacing between molecules of host-guest system \textbf{(a, c)} and naphtalene crystal \textbf{(b, d)}. The expansion is performed by first identifying individual molecules within the crystal and then increasing the distances between their centers of mass while keeping the molecular geometries and orientations rigidly fixed. The lowest-energy structure corresponds to the ground state of both PESs for a $2 \times 2 \times 3$ host-guest system containing 23 molecules (a, c) and for a $2 \times 2 \times 3$ naphtalene crystal containing 24 molecules. Differences between relative free energies predicted by reference DFT and G-MLIP$3$, and reference DFT and NMLIP are shown on the right axis (teal triangles). The effective volumes of C and H atoms used in calculating TS corrections are 32.25~\AA$^3$ and 7.57~\AA$^3$, respectively. The remaining parameters are obtained from the FHI-aims implementation of the TS vdW correction.}
\label{fig:SI_vdw_hostguest}
\end{figure*}

To assess the ability of MACE MLIPs to capture van der Waals (vdW) interactions, we calculated the relative energies of the host-guest system using G-MLIP3 upon cell expansion (Supplementary Figure \ref{fig:SI_vdw_hostguest}(a)). This was done by uniformly increasing all intermolecular distances while maintaining the molecular alignments. The range of mean intermolecular distances from 5.1~\AA~to~5.3~\AA~ corresponds to a $4\%$ increase in cell parameters, equivalent to a temperature increase from 80~K to 480~K. As shown in Supplementary Figure \ref{fig:SI_vdw_hostguest}(a), even a $1\%$ increase in intermolecular distances results in an energy error of approximately $20$~meV per molecule. G-MLIP3 exhibits similar errors for naphthalene molecular crystals.

To determine whether these errors originate from the potential itself, we tested NMLIP on a pure naphthalene crystal, for which we previously achieved the highest accuracies in energies, forces, and phonon frequencies. Supplementary Figure~\ref{fig:SI_vdw_hostguest}(b) shows the performance of NMLIP upon cell expansion up to $50\%$. While NMLIP performs significantly better than G-MLIP3 at short distances, it still fails to accurately capture long-range vdW interactions, which is naturally limited by the finite receptive field of MACE MLIPs.

To address long-range vdW interactions, we explicitly include vdW corrections using the Tkatchenko-Scheffler (vdW-TS) method~\cite{Tkatchenko2009}. Effective $C_6^\text{eff}$ coefficients were determined by using average Hirshfeld volumes of C and H atoms, which were obtained from DFT calculations over all structures in the training dataset and subsequently averaged. Since the original training data already contained dispersion corrections~\cite{Tkatchenko2012}, we subtracted the pairwise vdW-TS-like contributions from the reference energies and forces before training. The MACE potential was then trained on the residual interactions effectively removing vdW corrections. During prediction, the analytical vdW-TS model is added back to the energies, so that predicted values explicitly include long-range interactions.

Supplementary Figure~\ref{fig:SI_vdw_hostguest}(c) shows that, for the host-guest system, the explicitly vdW-corrected G-MLIP3 reduces short-range energy errors by a factor of two or more, keeping the training dataset unaltered.  Similarly, Supplementary Figure~\ref{fig:SI_vdw_hostguest}(d) shows that, for the pure naphthalene crystal, although small expansions are already well-described by the implicitly vdW-corrected NMLIP, the explicitly vdW-corrected NMLIP reduces long-range energy errors by approximately nine times.

%%%%%%%%%%%%%%%%%%%%%%55
%%%%%%%%%%%%%%%%%%%%%%%%%%%%%%%555555
%%%%%%%%%%%%%%%%%%%%%%%%%%%%%%%555555
\begin{figure*}[!h]
\centering
\includegraphics[width=18cm]{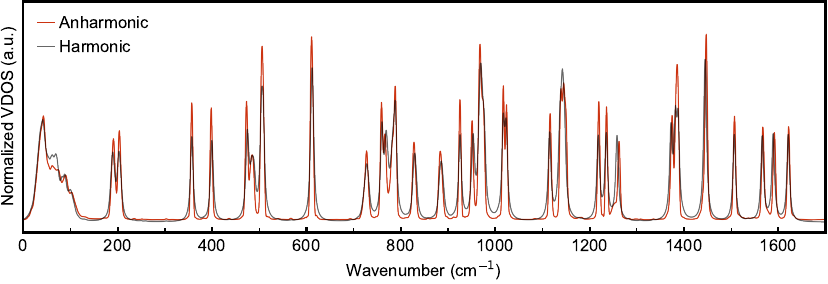}
\caption{\textbf{Comparison of harmonic and anharmonic VDOS of the host-guest system at $100$~K.} The harmonic VDOS is obtained by convolving the vibrational stick spectrum with a Lorentzian function of width $\wavenumber{3.3}$.}
\label{fig:SI_VDOS_AnharvsHar}
\end{figure*}
\begin{figure*}[!h]
\centering
\includegraphics[width=18cm]{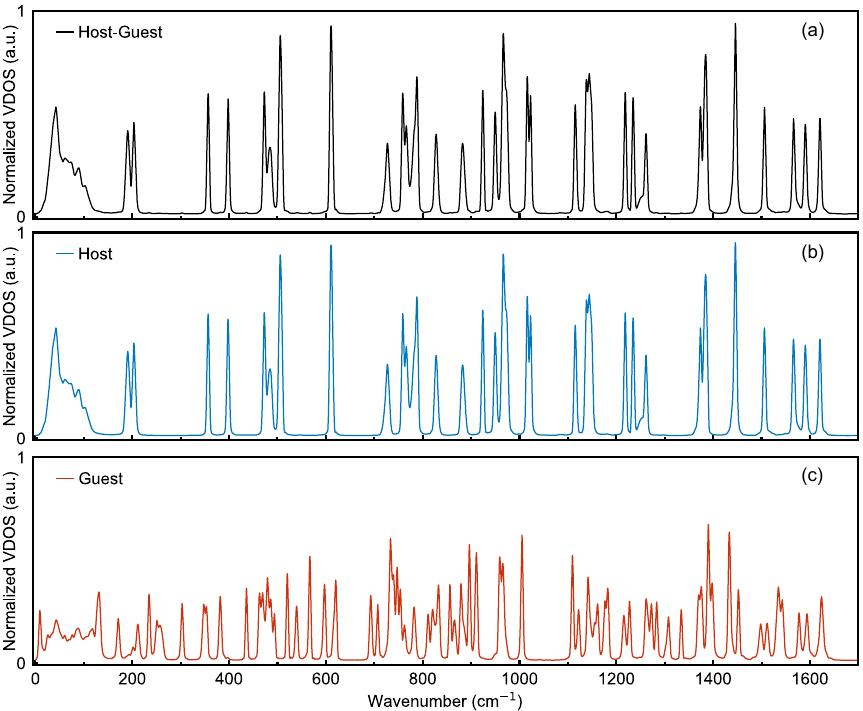}
\caption{\textbf{Comparison of projected VDOSs in the host-guest system.}~\textbf{a} host-guest, \textbf{(b)} host and \textbf{(c)} guest VDOSs. The guest VDOS is multiplied with $65$ for better visualization. All figures have the same scale. The green dashed lines denotes the vibrational frequencies of $78$ guest modes determined from the projected VDOS analysis shown in Figure ~\ref{fig:Fig_HG_VDOS}.}
\label{fig:SI_VDOS_individuala}
\end{figure*}

%%%%%%%%%%%%%%%%%%%%%%%%%%%%%%%555555
%%%%%%%%%%%%%%%%%%%%%%%%%%%%%%%555555
\begin{figure*}[!h]
\centering
\includegraphics[width=16cm]{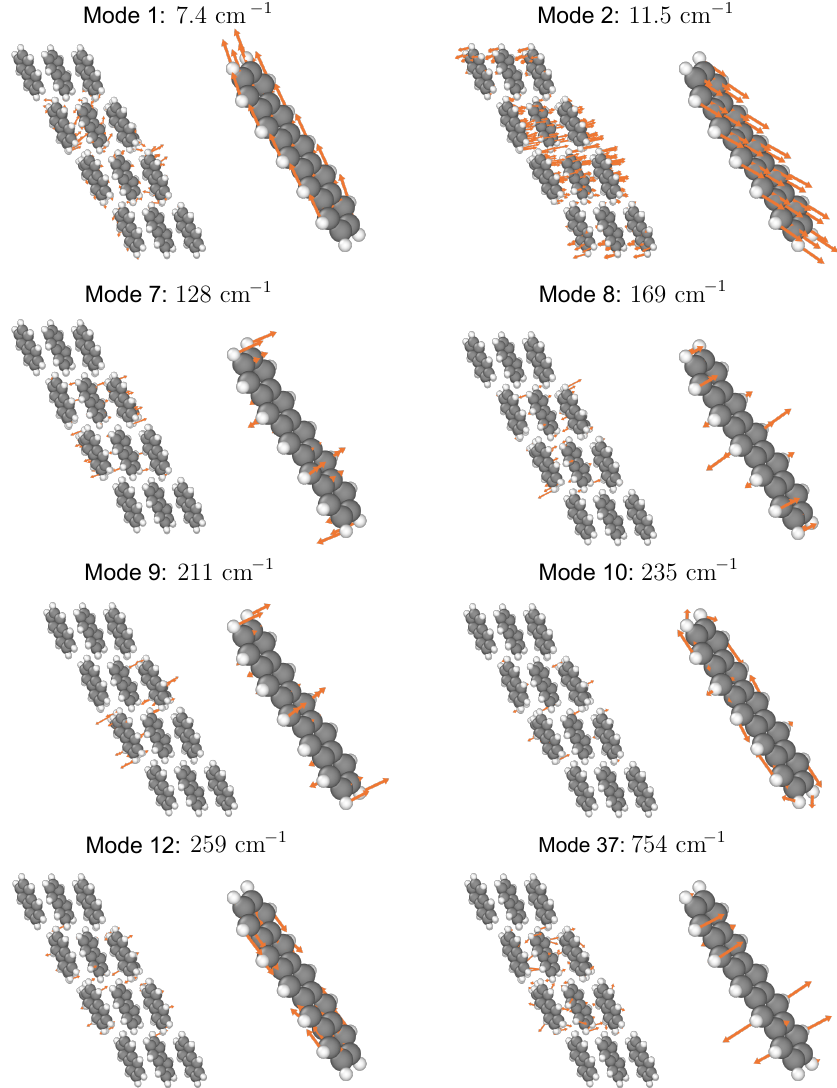}
\caption{\textbf{illustrations of mass-weighted host–guest vibrational modes shown in Fig.~\ref{fig:Fig_HG_VDOS}.} The left plots depict the modes projected onto a subset of host molecules, while the right plots show the corresponding modes on the guest molecule. The relative scaling between the host and guest eigenvectors is as follows: $4.5:1$ for modes $1, 7$, and $8$; $1.5:1$ for modes $2$ and $9$; $24:1$ for mode $10$; $48:1$ for mode $12$; and $36:1$ for mode $37$. These modes originate from vibrational modes of an isolated pentacene molecule in the gas phase at frequencies of $\wavenumber{36}, \wavenumber{36}, \wavenumber{100}, \wavenumber{148}, \wavenumber{191}, \wavenumber{239}, \wavenumber{265}$, and $\wavenumber{778}$, calculated at the B3LYP/def2-TZVP level of theory, respectively.}
\label{fig:SI_Host_Guest_Mode}
\end{figure*}
%%%%%%%%%%%%%%%%%%%%%%%%%%%%%%%555555

\begin{figure*}[!h]
\centering
\includegraphics[width=8cm]{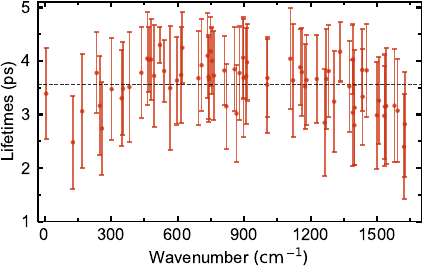}
\caption{\textbf{Vibrational lifetimes of the guest vibrational modes.} The lifetimes are extracted from the normal-mode projected VDOS of guest molecules as detailed in Methods.}
\label{fig:SI_HostGuest_uncertainity_lifetime}
\end{figure*}
%%%%%%%%%%%%%%%%%%%%%%%%%%%
\begin{figure*}[!h]
\centering
\includegraphics[width=16cm]{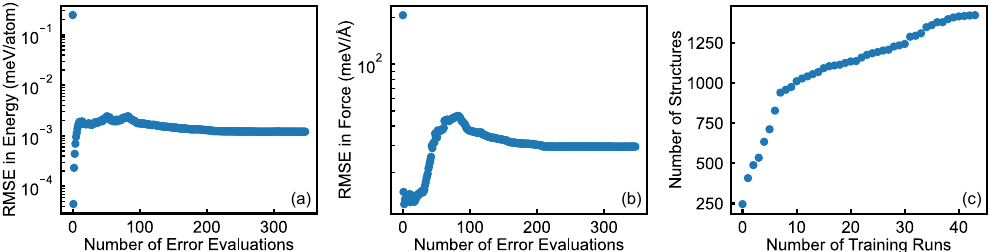}
\caption{\textbf{Metrics for VASP MLIP on-the-fly active-learning.} Training dataset RMSE errors in energy \textbf{(a)} and forces \textbf{(b)} for each error evaluation step. At each step, uncertain structures in a certain MD interval are added to the training dataset and MLIP is refitted.~\textbf{c} The number of structures in the training dataset during a $44$-step learning run, with each step lasting $5$~ps.}
\label{fig:SI_VASPTraining}
\end{figure*}
%%%%%%%%%%%%%%%%%%%%%%%%%%%%

\begin{figure*}[!h]
\centering
\includegraphics[width=17.8cm]{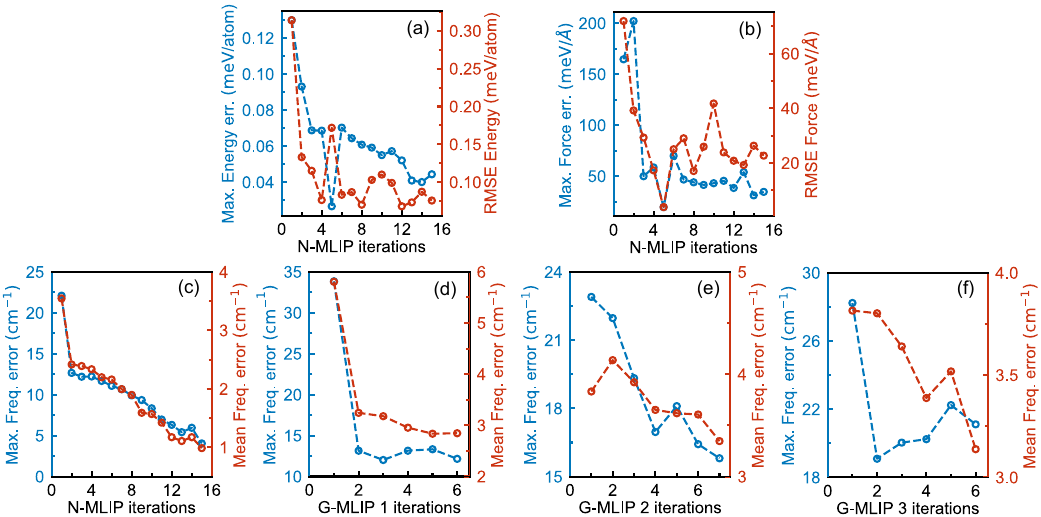}
\caption{\textbf{Evolution of errors during MLIP trainings.}~\textbf{a, b} Energy and force RMSE and maximum errors for the top 25 most uncertain structures in each N-MLIP iteration. \textbf{c-f} Maximum and mean errors in $\Gamma$-point phonon frequencies observed during the active-learning steps for N-MLIP~(c), G-MLIP1(d), G-MLIP2(e), and G-MLIP3(f).}
\label{fig:SI_TraniningErrors}
\end{figure*}
%%%%%%%%%%%%%%%%%%%%%%%%%%%

% \label{fig:SI_NMLP_training_energy_error}

%%%%%%%%%%%%%%%%%%%%%%%%5
\begin{table}[h] 
    \centering 
    \caption{The training and validation errors of committee mean for building N-MLIP. The errors are presented as RMSE for energies and forces.} 
    \label{tab:rmse_naph} 
    \begin{tabular}{ p{2cm} >{\centering\arraybackslash}p{3cm} >{\centering\arraybackslash}p{1.6cm} >{\centering\arraybackslash}p{1.6cm} >{\centering\arraybackslash}p{1.6cm} >{\centering\arraybackslash}p{1.6cm} >{\centering\arraybackslash}p{1.6cm} }
         & & \multicolumn{2}{c}{Energy (meV/atom)} & \multicolumn{2}{c}{Force (meV/\AA)} \\ 
        \hline
        AL-iteration & Number of training structures & Training & Validation & Training & Validation \\         \hline
        1  & 100 & 0.36 &0.39 & 3.07 &3.20 \\ 
        2  & 125 & 0.33 &0.38  & 2.20 &2.46  \\
        3  & 150 & 0.31 &0.33  & 2.11 &2.34  \\ 
        4  & 175 & 0.31 &0.33  & 2.01 &2.25  \\ 
        5  & 200 & 0.31 &0.33  & 1.93 &2.19  \\ 
        6  & 225 & 0.316 &0.316  & 1.889 &2.106  \\ 
        7  & 250 & 0.316 &0.316  & 1.843 &2.066  \\ 
        8  & 275 & 0.316 &0.316  & 1.829 &2.004  \\ 
        9  & 300 & 0.316 &0.316  & 1.763 &1.946  \\ 
        10  & 325 & 0.316 &0.316  & 1.748 &1.927  \\ 
        11  & 350 & 0.316 &0.316  & 1.720 &1.892  \\ 
        12  & 375 & 0.316 &0.316  & 1.694 &1.855  \\
        13  & 400 & 0.316 &0.276  & 1.672 &1.836  \\ 
        14  & 425 & 0.316 &0.276  & 1.667 &1.819  \\ 
        15  & 450 & 0.316 &0.316  & 1.614 &1.759  \\ 
        
        \hline
    \end{tabular}
\end{table}
%%%%%%%%%%%%%%%%%%%%%%%%%%555
\begin{table}[h] 
    \centering 
    \caption{The training and validation errors of committee mean for building G-MLIP1. The errors are presented as RMSE for energies and forces.} 
    \label{tab:rmse_anth} 
    \begin{tabular}{ p{2cm} >{\centering\arraybackslash}p{3cm} >{\centering\arraybackslash}p{1.6cm} >{\centering\arraybackslash}p{1.6cm} >{\centering\arraybackslash}p{1.6cm} >{\centering\arraybackslash}p{1.6cm} >{\centering\arraybackslash}p{1.6cm} }
         & & \multicolumn{2}{c}{Energy (meV/atom)} & \multicolumn{2}{c}{Force (meV/\AA)} \\ 
        \hline
        AL-iteration & Number of training structures & Training & Validation & Training & Validation \\         \hline
        1  & 475 & 0.316 &0.348 & 2.294 &2.445 \\ 
        2  & 500 & 0.508 &0.520  & 2.494 &2.713  \\
        3  & 525 & 0.510 &0.518  & 2.528 &2.832  \\ 
        4  & 550 & 0.508 &0.531  & 2.554 &2.897  \\ 
        5  & 575 & 0.545 &0.554  & 2.564 &2.925  \\ 
        \hline
    \end{tabular}
\end{table}
%%%%%%%%%%%%%%%%%5
\begin{table}[h] 
    \centering 
    \caption{The training and validation errors of committee mean for building G-MLIP2. The errors are presented as RMSE for energies and forces.} 
    \label{tab:rmse_tetra} 
    \begin{tabular}{ p{1.1cm} >{\centering\arraybackslash}p{2cm} >{\centering\arraybackslash}p{3cm} >{\centering\arraybackslash}p{1.6cm} >{\centering\arraybackslash}p{1.6cm} >{\centering\arraybackslash}p{1.6cm} >{\centering\arraybackslash}p{1.6cm} }
         & & \multicolumn{2}{c}{Energy (meV/atom)} & \multicolumn{2}{c}{Force (meV/\AA)} \\ 
        \hline
        AL-iteration & Number of training structures & Training & Validation & Training & Validation \\         \hline
        1  & 600 & 0.648 &0.648 & 2.895 &3.112 \\ 
        2  & 625 & 0.648 &0.658  & 2.876 &3.095  \\
        3  & 650 & 0.649 &0.651  & 2.865 &3.100  \\ 
        4  & 675 & 0.649 &0.649  & 2.834 &3.077  \\ 
        5  & 700 & 0.638 &0.651  & 2.832 &3.081  \\ 
        \hline
    \end{tabular}
\end{table}
%%%%%%%%%%%%%%%%%%%5
\begin{table}[h] 
    \centering 
    \caption{The training and validation errors of committee mean for building G-MLIP3. The errors are presented as RMSE for energies and forces.} 
    \label{tab:penta_rmse} 
    \begin{tabular}{ p{2cm} >{\centering\arraybackslash}p{3cm} >{\centering\arraybackslash}p{1.6cm} >{\centering\arraybackslash}p{1.6cm} >{\centering\arraybackslash}p{1.6cm} >{\centering\arraybackslash}p{1.6cm} >{\centering\arraybackslash}p{1.6cm} }
         & & \multicolumn{2}{c}{Energy (meV/atom)} & \multicolumn{2}{c}{Force (meV/\AA)} \\ 
        \hline
        AL-iteration & Number of training structures & Training & Validation & Training & Validation \\         \hline
        1  & 725 & 0.629 &0.640 & 2.794 &3.061 \\ 
        2  & 750 & 0.590 &0.620  & 2.794 &3.064  \\
        3  & 775 & 0.608 &0.620  & 2.805 &3.077  \\ 
        4  & 800 & 0.600 &0.620  & 2.799 &3.055  \\ 
        5  & 825 & 0.608 &0.631  & 2.794 &3.051  \\ 
        \hline
    \end{tabular}
\end{table}

% \bibliography{reference}% Produces the bibliography via BibTeX.
% \end{document}
%
% ****** End of file apssamp.tex ******

\end{document}